\documentclass{emulateapj}
\usepackage{apjfonts}

\newcommand{\Teff}{\mbox{$T_{\rm eff}$}}
\newcommand{\Porb}{\mbox{$P_{\rm orb}$}}
\newcommand{\Msun}{\mbox{$M_{\odot}$}}
\newcommand{\Mwd}{\mbox{$M_\mathrm{wd}$}}

\newcommand{\timav}{\langle \dot M \rangle}
\newcommand{\be}{\begin{eqnarray}}
\newcommand{\ee}{\end{eqnarray}}
\newcommand{\der}[2]{\frac{d #1}{d #2}}
\newcommand{\pder}[2]{\frac{\partial #1}{\partial #2}}
\shorttitle{CV WD $T_{\rm eff}$}

\begin{document}

%\submitted{Accepted by ApJL}
\title{Cataclysmic Variable Primary Effective Temperatures:\\
Constraints on Binary Angular Momentum Loss}

\author{Dean M. Townsley}
\affil{Department of Astronomy and Astrophysics,\\
University of Chicago, 5640 South Ellis Avenue, Chicago IL 60637;
townsley@uchicago.edu}

\and

\author{Boris T. G\"ansicke}
\affil{Department of Physics,\\
University of Warwick, Coventry CV47AL, UK; boris.gaensicke@warwick.ac.uk }

\begin{abstract}

We review the most decisive currently available measurements of the
surface effective temperatures, $T_{\rm eff}$, of white dwarf (WD)
primaries in cataclysmic variables (CVs) during accretion quiescence,
and use these as a diagnostic for their time averaged accretion
rate, $\timav$.  Using time-dependent calculations of the WD envelope,
we investigate the sensitivity of the quiescent $T_{\rm eff}$ to long
term variations in the accretion rate.  We find that the quiescent
$T_{\rm eff}$ provides one of the best available tests of predictions
for the angular momentum loss and resultant mass transfer rates which
govern the evolution of CVs.  While gravitational radiation is
completely sufficient to explain the $\timav$ of strongly magnetic CVs
at all $P_{\rm orb}$, faster angular momentum loss is required to
explain the temperatures of dwarf nova primaries (non-magnetic
systems).  This provides evidence that a normal stellar magnetic field
structure near the secondary, providing for wind launching and attachment,
is essential for the enhanced braking
mechanism to work, directly supporting the well-known stellar wind
braking hypothesis.  The contrast in $\timav$ is most prominent for
orbital periods $P_{\rm orb}> 3$ hours, above the so-called
\textit{period gap}, where $\timav$ differs by
orders of magnitude, but a modest enhancement is also present at
shorter $P_{\rm orb}$.  The averaging time which $\timav$ reflects
depends on $\timav$ itself, being as much as $10^5$ years for
low-$\timav$ systems and as little as $10^3$ years for high-$\timav$
systems.  We discuss in some detail the security of conclusions drawn
about the CV population in light of these time scales and our
necessarily incomplete sample of systems, finding that, due to the
time necessary for the quiescent $T_{\rm eff}$ to adjust, the
consistency of measurements between different systems places
significant constraints on possible long-timescale variation in $\dot
M$.  Measurements for non-magnetic systems above the period gap fall below
predictions from traditional stellar wind braking prescriptions, but above
more recent predictions with somewhat weaker angular momentum loss.
We also discuss the apparently high $T_{\rm eff}$'s found in the
VY Scl stars, showing that these most likely indicate $\timav$ in this
subclass even larger than predicted by stellar wind braking.

\end{abstract}


\defcitealias{TownBild04}{TB04}

%%%%%%%%%%%%%%%%%%%%%%%%%%%%%%%%%%%%%%%%%%%%%%%%%%%%%%%%%%%%%%%%%%%%%%%%
\section{Introduction}

The evolution of short-period binaries containing a Roche lobe filling
low-mass main sequence (MS) star transferring matter onto a white
dwarf (WD), the bulk of the cataclysmic variables (CVs;
\citealt{Warn95}), has long been believed to be driven by two distinct
angular momentum loss mechanisms.  Each of these has a distinct rate
at which angular momentum is lost, $\dot J$, and resulting
time-averaged mass transfer rate $\timav$ required such that the MS star
remains within the Roche lobe dimensions.  The absolute minimum
$\dot J$ for the binary is set by losses due to gravitational
radiation from the orbital motion itself
\citep{Faul71,PaczSien81,Rappetal82}.  Higher $\dot J$ is obtained due
to the partial magnetic attachment of stellar winds similar to those
which spin down isolated stars \citep{VerbZwaa81}, which extract
angular momentum from the orbit because the Roche lobe filling MS star is
tidally locked to the orbit.  It was suggested
\citep{PaczSien83,Rappetal83,SpruRitt83} that the evolution of CVs
could be explained by combining these two mechanisms: at long orbital
periods, $P_{\rm orb}\gtrsim 3$~hours, when the star has mass $M_{\rm
MS}\gtrsim 0.25M_\odot$, and can support a typical stellar
magnetosphere, wind losses dominate $\dot J$, but when $M_{\rm
MS}\lesssim 0.25M_\odot$, the star's magnetic structure is disrupted
by the loss of the radiative core so that $\dot J$ falls to the level
of gravitational radiation.

This interrupted magnetic braking (IMB) scenario \citep[see
e.g.][]{Hameetal88,Kolb93,Howeetal01} explains an essential feature of
the observed CV population: the lack of systems with $2~{\rm hours}
\lesssim P_{\rm orb} \lesssim 3$~hours.  Systems in this range are
predicted to be detached and non-mass transferring because the
high mass loss rate to which the MS star was subject during the wind $\dot
J$ phase is large enough for it to become bloated, such that when
$\dot J$ decreases and it returns to the equilibrium radius for a main
sequence star of the appropriate mass, it is well within the Roche
lobe.  The orbit must then contract further, to shorter $P_{\rm orb}$,
before contact and thus mass transfer is reestablished.  Such bloating
above the period gap does appear to be borne out by recent studies
\citep{beuermannetal98-1, knigge06-1}.  While the relation
derived from the spindown of cluster stars can, with questionable
justification, be extrapolated to the spins appropriate to the CV $P_{\rm
orb}$ values, what has actually taken place in practice is that several
variations were considered early on \citep{Rappetal83} and that which best
reproduced the observed period gap was chosen.  Thus the mass transfer rate,
$\dot M$, which IMB predicts above the period gap is that which is necessary
to bloat the MS star by the necessary amount to reproduce the $P_{\rm orb}$
extent of the gap itself.

While the IMB scenario of CV evolution does explain the observed
period gap, it faces a number of problems. On one hand, observations
of single low-mass stars do not show evidence for a change in
spin-down rate at the mass boundary where the stars become fully
convective \citep{Andretal03}. On the other hand, most attempts to
observationally measure $\dot M$ in both $\dot J$ regimes have proven
difficult.  Initial indicators from disk properties
\citep{patterson84-1} were consistent with the IMB picture, but they
lack accuracy as they rely on a number of poorly known system
properties, including their distances. Even if successful, $\dot M$
determined from accretion disk studies measures only the instantaneous
mass transfer rate.  Indicators for $\timav$ from the departure from
thermal equilibrium in the donor star appear quite promising, but
suffer from the fact that reproducing the radii of isolated stars
with these theoretical models is non-trivial \citep{Beue06,Riba06},
and the value of $\dot M$ derived can be very sensitive to the radius
predicted by the model.  \citet{TownBild05} were able to use the
period-specific classical nova rate to investigate the $\dot
M$-$P_{\rm orb}$ relation since the nova ignition mass, and therefore
inter-outburst time, depends on $\timav$.  However absolute $\dot M$
measurements with this method are elusive due to uncertainties in the
mass distribution and CV population density.  Measurements of the
WD effective temperature, $T_{\rm eff}$, the subject of this paper,
provide one of the more promising avenues for constraining the $\dot
M$ in observed systems.  The quiescent $T_{\rm eff}$ expected from a
given time average accretion rate $\timav$ can be calculated directly
(\citealt{TownBild03}; \citealt{TownBild04}; herafter TB03 and TB04),
and our discussion will largely be concerned with how representative
the $\timav$ inferred $T_{\rm eff}$ is of the true $\timav$.

Given that the WDs in CVs are relatively hot objects,
$\Teff\ga10\,000$\,K, their spectral energy distribution peaks in the
ultraviolet (UV), and it is in this wavelength range that most of the
CVWD \Teff\ measurements were obtained. A handful of bright CVWDs were
intensively studies with the \textit{International Ultraviolet
Explorer} (\textit{IUE}), e.g. VW\,Hyi \citep{mateo+szkody84-1},
WZ\,Sge \citep{sionetal90-1}, or AM\,Her \citep{heise+verbunt88-1,
gaensickeetal95-1}. Temperature estimates obtained prior to the launch
\textit{Hubble Space Telescope} (\textit{HST}) were summarized by
\citet{Sion91}, and updated for measurements obtained predominantly
with the \textit{HST} first-generation UV spectrographs by
\citet{Sion99}. A significant increase in the number of reliable CVWD
\Teff\ measurements has become available since the deployment of the
Space Telescope Imaging Spectrograph (STIS) on \textit{HST} and the
launch of the \textit{Far Ultraviolet Spectroscopic Explorer}
(\textit{FUSE}), see e.g. \citet{araujo-betancoretal05-2}.

We begin by reviewing the relationship between $\timav$ and $T_{\rm
eff}$ as discussed by TB03.  The amount of variation expected in
$T_{\rm eff}$ due to long term $\dot M$ variations is evaluated from
both quasi-static models and time-dependent envelope simulations.
Following this, in \S\ref{sec:system}, features of the system
such as outburst properties, accretion geometry, and possible
additional heat sources are discussed to evaluate their impact on the
utility of $T_{\rm eff}$ as and indicator of $\timav$.
\S\ref{sec:measurements} critically reviews the available measurements
of CV WD $T_{\rm eff}$'s, with the intention of selecting a
well-understood set of measurements from which firm conclusions on CV
properties can be drawn.  After some discussion of general
uncertainties and biases, \S\ref{sec:discussion} sets out
our conclusions, including the contrast in $\timav$ across the
gap, the enhanced $\timav$ in VY\,Scl-type novalikes displaying
low states, and the case for the suppression of wind braking in CVs
with highly magnetic WDs.

%%%%%%%%%%%%%%%%%%%%%%%%%%%%%%%%%%%%%%%%%%%%%%%%%%%%%%%%%%%%%%%%%%
\section{Relationship of $T_{\rm eff}$ to $\dot M$}
\label{sec:theory}

The luminosity streaming up through the surface of the WD during accretion
quiescence is released at all depths effectively down to the base of the
accreted layer by the compression of material as it is pushed deeper into the
star by further accretion (TB04).  The infall energy is deposited at very
shallow depths and therefore radiates away
quickly ($\sim$ hours) after the cessation of active accretion (see
\S\ref{sec:reachingquiescent} below for more discussion).
Due to the lengthening thermal time with increased depth, a large
portion of this quiescent luminosity reflects an accretion history averaged
over timescales which can be longer than $10^4$ yr, depending on the
characteristic value of $\dot M$.

As a reference point from which to begin, \S\ref{sec:averageTeff}
reviews how the quiescent luminosity depends on the time averaged value of
the accretion rate, $\timav$, and other features of the WD.  This provides a
relation between $T_{\rm eff}$ and $\timav$ which has no ``free'' parameters,
but does have important but manageable uncertainties related to unknown
properties of the WD.  Following this, in \S\ref{sec:timescales} we present a
basic discussion of how the response of the envelope to accretion can be
understood in terms of the run of local thermal time with depth.  This
provides justification for the general assertions made above about the
timescales on which the surface luminosity can vary.  Using estimates of the
run of thermal timescale in the outermost portion of the envelope, in
\S\ref{sec:reachingquiescent} we derive the timescales on which heat
deposited near the surface is radiated away.  This demonstrates how
measurement of the luminosity escaping due to compression is possible between
transient accretion events such as dwarf novae.  Such simple estimates are
insufficient for characterizing long-term variations, so that we proceed, in
\S\ref{sec:simulations}, to discuss and then characterize, with simulations,
how long timescale variations in $\dot M$ affect the quiescent $T_{\rm eff}$.

\subsection{Long Term Average Properties}
\label{sec:averageTeff}

\citetalias{TownBild04} presented a detailed discussion of the impact
of accretion at rates low enough that hydrogen burns in unstable
outbursts, Classical Nova eruptions, on the thermal
structure of a WD.  For this paper we are primarily interested in the
predictions for the quiescent surface luminosity, $L_{\rm
q}(M,\timav,T_c)$, where $M$ is the WD mass, $\timav$ is the
time-average accretion rate, and $T_c$ is the WD core temperature.
Through a simple balance of compression of material and radiative heat
transport, the quiescent surface luminosity is given by $L_{\rm
q}\simeq \timav T_b/\mu m_p= 4\pi R^2 \sigma T_{\rm eff}^4$ where
$T_b$ is the temperature at the base of the radiative layers,
typically $T_b\simeq T_c$, $R$ is the WD radius, $\mu$ is the mean
molecular weight of the accreted material, $m_p$ is the mass of the
proton and $\sigma$ is the Stefan-Boltzmann constant.  TB04 found that
$T_c$ approaches an equilibrium value when the WD is subject to
accretion at constant $\timav$ for timescales similar to the WD core
thermal time, $\sim 10^8$ years.  This $T_{c,\rm eq}$ is not strongly
sensitive to $M$ and increases weakly with $\timav$, being generally
0.5 to 1 $\times 10^7$ K for CV WDs, which have $\timav = 10^{-11}$ to
$10^{-8}M_\odot\ {\rm yr^{-1}}$.  This equilibrium has recently been
demonstrated in simulations which follow the WD evolution through many
nova outbursts, finding slightly but not significantly higher
$T_{c,\rm eq}$ for the higher $\timav$'s in this range, and similar
evolutionary timescales \citep{Epeletal07}.  $L_{\rm q}$ is
insensitive to $T_c$ for $T_c\le T_{c,\rm eq}$ and the $\timav$
relevant above the period gap, where $\timav$ changes on timescales
less than the core thermal time (TB03).  Thus we do not need to know
the details of $T_c$ in order to derive $\timav$ from $T_{\rm eff}$,
though we use $T_{c,\rm eq}$ as a convenient reference value.

Including a small amount of nuclear heating, using the methods of TB04
we find that (at $T_{c,\rm eq}$) the average $L_{\rm q}$ during the
classical nova (CN) cycle is
\begin{equation}
\label{eq:lumrelation}
L_{\rm q} = 6\times 10^{-3} L_\odot
\left(\frac{\timav}{10^{-10}M_\odot\ {\rm yr^{-1}}}\right)
\left(\frac{M}{0.9M_\odot}\right)^{0.4}.
%L/L_\odot = 1.8142\times 10^8
%left(\frac{\timav}{M_\odot\ {\rm yr^{-1}}}\right)^{1.056}
\end{equation}
Using an approximate power-law relation for $R\propto M^{-1.8}$ near
$M=1.0M_\odot$, this gives
\begin{equation}
\label{eq:Teffrelation}
T_{\rm eff} = 1.7\times 10^4\ {\rm K}
\left(\frac{\timav}{10^{-10}M_\odot\ {\rm yr^{-1}}}\right)^{1/4}
\left(\frac{M}{0.9M_\odot}\right)\ .
%\left(\frac{M}{0.6M_\odot}\right)^{0.975}.
\end{equation}
As emphasized in TB03, the $\timav$ inferred from a system's quiescent
$T_{\rm eff}$ is strongly dependent on the assumed $M$, almost entirely due
to the use of $R$ to infer $L_{\rm q}$.  While $\timav\propto M^{-3.9}$ for a given
$T_{\rm eff}$, $\langle\dot m\rangle \equiv \timav/4\pi R^2\propto M^{-0.3}$.
Just a 25\% uncertainty in mass, allowing $M=0.6$ to $1.0M_\odot$, leads to
nearly a \emph{factor of ten} uncertainty in  $\timav$.  This seems to be the
strongest limit on the
utility of measurements of $T_{\rm eff}$, and is difficult to avoid without
either independent mass measurements or distances.
Only a handful of systems have independent mass measurements (see Table
\ref{t-teff}), and all but one are DN below the period gap, and the one above
is a VY Scl star with a high $\timav$.  It appears infeasible to make
conclusions based on a subsample that only includes objects with mass
measurements.
However, as we will see
below, order of magnitude contrasts are easily discernible when the full
sample is considered and are
important for discriminating between angular momentum loss laws.
Additionally,  there are now enough sound measurements that further progress
can be made with some assumptions about the mass distribution of the
population.

The next most important source of uncertainty, which is nearly
impossible to eliminate for most systems, is due to the increase in
$L_{\rm q}$ as the mass of the accumulated layer, $M_{\rm acc}$,
increases between classical nova outbursts.  Scatter in the
observed $T_{\rm eff}$ values at the level of $\pm5\%$ is expected due
to systems having different $M_{\rm acc}$'s at the current epoch.
This degree of scatter is drawn from allowing a range $0.05M_{\rm
ign}\le M_{\rm acc} \le 0.95 M_{\rm ign}$, and, though the full allowed
range gives a larger variation, this represents well the size of layer
an observed WD is likely have. (See TB04 for details of how $T_{\rm
eff}$ varies with $M_{\rm acc}$.) We set the mass fraction $X_{^3\rm
He}=0.001$ in the accreted material throughout this work, because the
difference between this and similar predictions for $X_{^3\rm
He}=0.005$ is less than the uncertainty due to the unknown $M_{\rm
acc}$ (TB04).

\subsection{Envelope Response Timescales}
\label{sec:timescales}

One of the attractions of utilizing $T_{\rm eff}$ as an indicator of $\timav$
is that $T_{\rm eff}$ contains averaged information about the time history of
$\dot M$ rather than its instantaneous value.  This is of great benefit for
comparing $\timav$ with predictions based on long-term drivers of the binary
evolution contributing to orbital angular momentum loss.  Before proceeding
with simulations of the response of the envelope to $\timav$ variations, it
is useful to first discuss the relevant timescales within the envelope and
how they relate to the surface luminosity.  We do this in order to understand
how much a perturbation on some prescribed timescale is likely to affect the
surface luminosity.

The heat equation in the outer layers of the WD is given by (TB04)
\begin{equation}
\label{eq:heat}
 c_P \pder{T}{t} = \pder{F}{y} +c_P \dot m\frac{T}{y}(\nabla-\nabla_{\rm
ad}),
\end{equation}
Where $y = \int_r^R\rho dr\simeq P/g$, the column depth in from the
surface ($r=R$), forms a radial coordinate, $P$ is the pressure at radius $r$,
$g$ is the surface gravity, $F=
(4\sigma T^3/3\kappa)dT/dy$ is the local area-specific energy flux with
$\sigma$ being the Stefan-Boltzmann constant, $\kappa$
the opacity, $c_P$ the specific heat at constant pressure, $\nabla
= d\ln T/d\ln P$, $\nabla_{\rm ad} = \partial \ln T /\partial \ln P$ at
constant entropy, and $\dot m = \dot M/4\pi R^2$.  The temperature
profile of the envelope at a given depth, $y$, can change on the
thermal time for that layer, which we define by ``one-zone''
differencing the left hand side and the first right hand term in
equation (\ref{eq:heat}), dropping the accretion source term, to
obtain,
\begin{equation}
\tau_{\rm th} \equiv \frac{yc_P T}{F} \simeq \frac{3\kappa y^2c_P T}{4\sigma
T^4}.
%\tau_{\rm th} \equiv \frac{4\sigma T^3}{\kappa}\pder{T}{y},
\end{equation}
In a static envelope state, which is a good approximation for a steady
$\dot M$ (TB04), $\partial T/\partial t$ is small or zero, so that
$\partial F/\partial y$, the contribution to the surface flux from
each layer, is set by $\dot m$ and local properties of the layer.
Thus we expect that variations in $\dot M$ will appear as variations
in $\partial F/\partial y|_y$ after a time $t_{\rm th}(y)$, when that
layer can respond.  For order unity variations in $\dot M$ on a given
timescale $t_{\rm var}$, the contribution to $L_{\rm q}$ from layers
which have $t_{\rm th}>t_{\rm var}$ will not be affected by the
variation, while the contribution from layers with $t_{\rm th}<t_{\rm
var}$ will change by unity along with $\dot M$.  Thus $L({\rm surf})
-L(t_{\rm th}=t_{\rm var})$, the contribution to the surface
luminosity from the layers outside that with $t_{\rm th} = t_{\rm
var}$, evaluated with the average envelope state, provides a reasonable
indication of the possible variation in $L_{\rm q}$ which can be expected due
to unity-level variations in
$\dot M$ on a timescale of $t_{\rm var}$.  This is very similar to how
the characteristic cooling time of a dwarf nova outburst is set by the
thermal time at the bottom of the freshly accreted material
\citep{piroetal05-1}.

This analysis no longer applies in the deepest layers where the
thermal transport becomes mediated by electron conductivity.  This is
because the one-zoning used to define $\tau_{\rm th}$ no longer
holds.  Once $t_{\rm var}$ exceeds the thermal time of the whole
radiative layer, we are left with a problem which is more similar to
the classic WD cooling problem: an insulating layer which acts as the
thermal regulator for the underlying heat reservoir formed by
layers which are much more thermally well-coupled by electron
conduction.  In this case the timescale for further changing $L_{\rm
q}$ is expected to approach the CN inter-outburst time, that required
to build up the maximum degenerate region.

\subsection{Reaching the quiescent $T_{\rm eff}$}
\label{sec:reachingquiescent}

With the above estimate for thermal response time with depth, it is useful at
this point to discuss specifically the cooling of the thin outer layer that
is heated by the infall energy of the accreted matter.  This is essential to
understanding why the quiescent $T_{\rm eff}$ is a good indicator of the
energy being liberated by compression in the deeper layers of the star.
When material reaches
the surface of a WD via the disk, half of the gravitational infall
energy ($GM/R$ per unit mass, where $G$ is Newton's constant) has been
radiated in the disk and the rest is possessed as kinetic energy.  This
kinetic energy is deposited immediately at the WD surface as the
material is stopped and spread by interaction with the surface layers
\citep{piro+bildsten04-1}.  This process can heat the WD surface to
quite high temperatures given by $T_{\rm surf} = [(GM\dot M/2R)/4\pi
R^2\sigma]^{1/4} \approx 10^5~{\rm K}\, (\dot M/10^{-8} M_\odot~{\rm
yr}^{-1})^{1/4}(M/0.9M_\odot)^{1.6}$. 

However, this heat does not
penetrate the WD due to the opposing thermal gradient in the
underlying radiative atmosphere which has a profile characterized by
$F_q = L_q/4\pi R^2$ satisfying $4\sigma T^4/3\kappa y =
F_q$.  During accretion, the infall heating can penetrate
to approximately where the local temperature is the same as $T_{\rm
surf}$, or a column depth of $ y_{\rm surf}\simeq 4\sigma T_{\rm surf}^4/3\kappa
F_{\rm q}$. Thus, if $\kappa$ is assumed to be dominated by
electron scattering at this depth for simplicity,
the mass of the layer heated by infall is $M_{\rm surf} = 4\pi R^2y_{\rm
surf} \simeq 5\times 10^{-11}M_\odot(T_{\rm surf}/10^5{\,\rm
K})^{4}(T_{\rm eff,q}/14\,{\rm kK})^{-4}$.
The thermal
time for this layer to cool is between $ yc_PT/F_{\rm surf}$ and
$yc_PT/F_{\rm q}$ which are $10^2$ and $10^5$ seconds for $T_{\rm
eff,q}=14$ kK.  The latter is fairly consistent with the 2.8 days found for
the cooling of VW Hyi following a normal outburst
\citep{gaensicke+beuermann96-3}.
This timescale should be typical for cooling after a normal
dwarf nova outburst, but is much shorter than the cooling time after a
superoutburst, in which an order of magnitude more material than this is
deposited.  In that case, in contrast, heat is released on the cooling time
for the added material due to its relatively rapid compression
\citep{piroetal05-1}.  Note that in reality much of the material added to the
WD cools as it spreads over the surface \citep{piro+bildsten04-1} so that the
above analysis only applies to the heated region near the equator while the
rest of the star remains near $T_{\rm eff,q}$.

\subsection{Numerical Simulations of Long Term Accretion Rate Variation}
\label{sec:simulations}

The analysis in \S\ref{sec:timescales} indicates that a given variation in
$\dot M$ leads to a larger variation in the quiescent surface flux, $L_{\rm
q}$, if it occurs on a longer timescale.  That is, the excursion from the
average quiescent luminosity $\delta L_{\rm q} = \max|L_{\rm q} - \langle
L_{\rm q}\rangle|\approx (L_{\rm q,max}-L_{\rm q,min})/2$ increases with the
timescale of $\dot M$ variation, $t_{\rm var}$, for a fixed
amplitude of $\dot M$ variation.
Thus very brief variations in $\dot M$, such as dwarf novae outbursts,
have very little impact on $L_{\rm q}$ despite their large magnitude in $\dot
M$, once the shallow transient has passed.  The luminosity excursion, $\delta
L_{\rm q}$, of course also depends on the magnitude of the $\dot M$
variation.  This dependence is expected to be fairly simple, roughly linear,
and therefore we focus here on the dependence on $t_{\rm var}$, which relates
to the thermal response time structure of the envelope.

In order to test how $L_{\rm q}$, and
therefore the quiescent $T_{\rm eff}$, will vary in response to $\dot
M$ variations, we have performed simulations of an accreting layer on
a WD in the plane-parallel approximation.  This approximation is fairly
good through the accreted layer and has little impact on the
time-variation properties being studied here.  A similar
approximation, extending only to more shallow depths, was used by
\citet{piroetal05-1} in studying the decline of $T_{\rm eff}$ after
a dwarf nova superoutburst. Our numerical treatment is described in the
appendix.

As simple experiments, we have applied two forms of
long-term $\dot M$ variation: sinusoidal, with $\dot M = \timav
(1-0.8\sin[\pi t/P_{\rm 1/2}])$, and square
wave, with $\dot M = \timav (1-0.8\;\textrm{sign}[\sin(\pi t/P_{\rm
1/2})])$, for a wide range of half-period of $\dot M$ variation,
$P_{\rm 1/2}$.  Short term variations in $\dot M$ such as accretion
disk cycles (dwarf nova outbursts) are not treated explicitly in order
to keep timesteps large.
\begin{figure*}
%\plotone{L-time_combine.eps}
\plotone{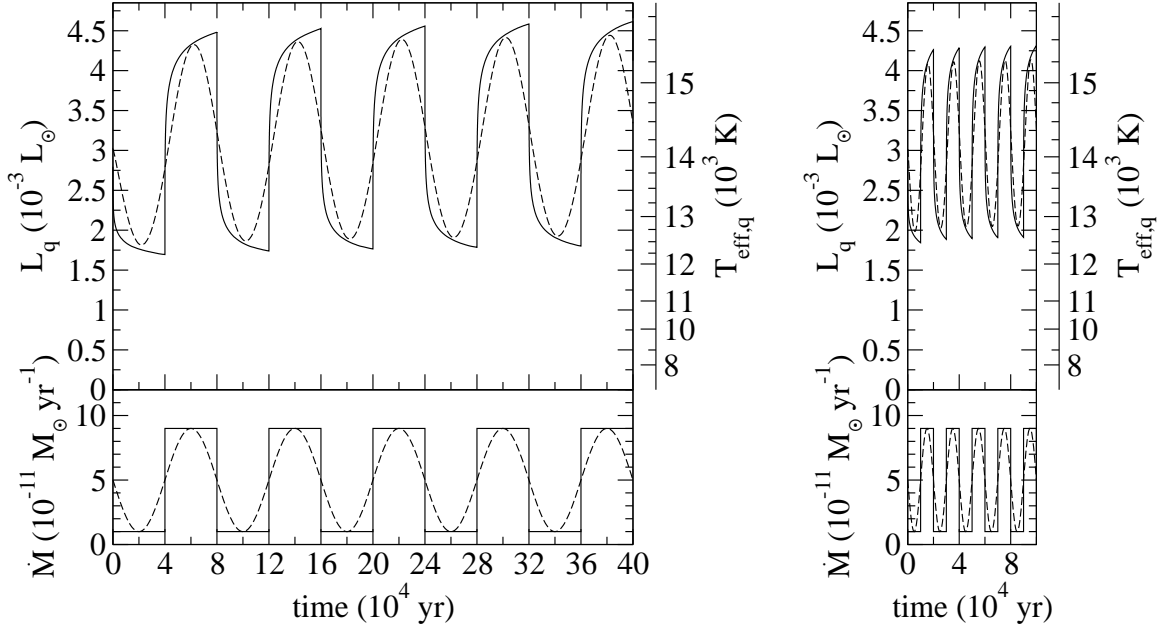}
%\plotone{L-time.eps}
\caption{
\label{fig:L-time}
Time evolution of quiescent surface flux, $L_{\rm q} = 4\pi R^2\sigma T_{\rm
eff}^4$, due to compressional energy release in the envelope for a WD with
$M=0.9M_\odot$ and $\timav = 5\times 10^{-11}M_\odot$ yr$^{-1}$.  Two
variations are applied, a sine (dashed lines) and a square wave (solid lines)
both with amplitude of $0.8\timav$.  Two timescales of applied variability are
shown, $P_{1/2}=10^5$ (left) and $4\times10^4$ yr (right).
For the same magnitude variation in $\timav$, a longer timescale of
variation leads to a slightly larger variation in $T_{\rm eff}$.
}
\end{figure*}
Figure \ref{fig:L-time} shows time series of $L_{\rm q}$
and $T_{\rm eff}$ for variations in $\dot M$ on two timescales,
$P_{1/2}=10^5$ and $4\times 10^{4}$ yr.  This example uses $M=0.9M_\odot$ and
$\timav = 5\times 10^{-11}M_\odot$ yr$^{-1}$, representative of DN systems
below the period gap, and has $T_{\rm eff,q}=14$ kK and $M_{\rm
ign}=2.3\times10^{-4}M_\odot$.  Time zero corresponds to $M_{\rm acc} = 0.5
M_{\rm ign}$.  The model is initialized by starting from the static solution
described in TB04 with $M_{\rm acc} = 0.25 M_{\rm ign}$ and applying $\dot
M=\timav$ until $M_{\rm acc}=0.5M_{\rm ign}$ to allow the model to settle on
the new numerical grid.  In Figure~\ref{fig:L-time}, a modest increase in the
total variation is observed with the square wave $\dot M$ applied instead of
the sinusoid.  There is no appreciable phase difference between the
applied $\dot M$ variation and that the response.  As expected, for the same
$\timav$ and magnitude of $\timav$ variation, a longer timescale leads to a
larger variation in $T_{\rm eff}$.  Consecutive peaks are slightly increasing
due to the increasing $M_{\rm acc}$ during the buildup to classical nova, the
longer displayed example making it to nearly $M_{\rm acc} = 0.59M_{\rm ign}$.

\begin{figure}
%\plotone{var-time.eps}
\plotone{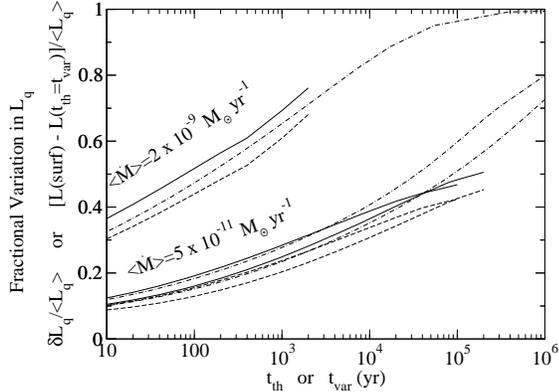}
\caption{
\label{fig:var-time}
Fractional excursion in quiescent surface flux, $\delta L_{\rm q}/\langle
L_{\rm q}\rangle = (L_{\rm q,max}-L_{\rm q,min})/2\langle L_{\rm q}\rangle$,
as a
function of timescale of the applied $\dot M$ variability, $t_{\rm var}$.
The response of an envelope simulation to a $0.8\timav$ variation in
$\dot M$ is shown for a square (solid) and sinusoidal (dashed) variation.
One mass is shown for $\timav = 2\times
10^{-9}M_\odot$~yr$^{-1}$, with $M=0.9M_\odot$, (far upper) and two for
$\timav = 5\times 10^{-11}M_\odot$ yr$^{-1}$, with $M=0.9$ (middle) and
$0.6M_\odot$ (lower).  These cases have $\langle L_{\rm q}\rangle$ ($\langle
T_{\rm eff}\rangle$) of 120, 2.8, and 2.6 $\times 10^{-3}L_\odot$ (35, 14,
and 12 kK) respectively.  Also shown for comparison (dot-dash lines) is the
run of $[L({\rm surf})-L(t_{\rm th}=t_{\rm var})]/\langle L_{\rm
q}\rangle$ in a static approximation of the average envelope state.  This
represents at what depth, in terms of timescale, energy is being released by
compression,
and provides a good estimate of the variation expected in response to order
unity $\dot M$ variation except at the deepest parts of the envelope where
electron conduction becomes important.
}
\end{figure}

There is a tremendous contrast between observable timescales of tens of
years, and the time between classical nova outbursts which can be as much as
$10^8$ years depending on $\timav$.  In order to probe this variety,
we have performed the same 5-cycle simulations shown in Figure
\ref{fig:L-time} for variability timescales up to about 2.5\% of the classical
nova accumulation time, making the total simulation time about 25\% of the
accumulation time.  We characterize the variation in $L_{\rm q}$ by its
excursion $\delta L_{\rm q}\approx \langle L_{\rm q,max} - L_{\rm
q,min}\rangle/2$, where the maximum and minimum are evaluated for each cycle
and the difference is then averaged over all cycles.  This is then divided by
the mean, $\langle L_{\rm q}\rangle$, to obtain a fractional variation.
Figure \ref{fig:var-time} shows the excursion found from the numerical
simulations using square (solid lines) and sine (dashed lines) wave $\dot M$.
Three cases are shown, $M=0.9M_\odot$ with $\timav = 2\times
10^{-9}M_\odot$~yr$^{-1}$ and $M=0.9$ and $0.6M_\odot$ with $\timav = 5\times
10^{-11}M_\odot$ yr$^{-1}$.  These have $\langle L_{\rm q}\rangle$ ($T_{\rm
eff}$) of 120, 2.8, and 2.6 $\times 10^{-3}L_\odot$ (35, 14, and 12 kK)
respectively, $M_{\rm ign}$ of 2.2, 23 and 39$\times 10^{-5}M_\odot$, time
between classical novae of $1.1\times 10^4$, $4.6\times 10^6$ and $7.8\times
10^6$ yr, and $T_c$ of 9.6, 5.7 and $5.7\times 10^6$ K.

As discussed in \S\ref{sec:timescales}, the magnitude of the variation or
excursion in $L_{\rm q}$ can be estimated by evaluating the compressional
heat release at depths with thermal time less than $t_{\rm var}$.  This can
be expressed as $L({\rm surf})-L(t_{\rm th}=t_{\rm var})$, where the
quiescent surface luminosity, $L(\rm surf)$ and the luminosity at the depth
where the thermal time matches the variation time of the applied $\dot M$
variation, $L(t_{\rm th}=t_{\rm var})$ are evaluated from the run of $L$ and
$t_{\rm th}$ in a static accreting envelope at $\dot M = \timav$, $L(\rm
surf) = \langle L_{\rm q}\rangle$ and $M_{\rm acc} = 0.5 M_{\rm ign}$.  These
curves are shown for comparison in Figure \ref{fig:var-time}, one for each
case shown from the time-dependent simulations.

We find that $L({\rm surf})-L(t_{\rm th}=t_{\rm var})$
provides a good estimate of the variability until
electron conduction becomes important, $t_{\rm th}\gtrsim 10^4$ yr.
Deeper than this, the $t_{\rm th}$ we have defined underestimates the
thermal time, so that $L(t_{\rm th}=t_{\rm var})$ is being
evaluated at too deep a layer and $L({\rm surf})-L(t_{\rm th}=t_{\rm var})$
is therefore being overestimated.  The variability seen in $L_{\rm q}$ in the
simulations does not
increase for longer timescales, due to the increased heat capacity
from the thermally connected layers where electron conduction provides
heat transport that is efficient relative to the overlying radiative
layer.  We find that the averaging time for a given system depends strongly
on its $\timav$.  This is due to the fact that while the luminosity varies by
two orders of magnitude (approximately  $\propto\timav$) for CVs, the thermal
content of the radiative layers of the envelope changes little with $\timav$.
Note that we have only explored order unity $\dot M$ variations here for
simplicity of demonstration, smaller variations will lead to correspondingly
smaller responses.

The magnitude of the variation in quiescent surface luminosity, $L_{\rm q}$,
depends on the timescale of the proposed variation in $\dot M$.  Since the
the $\timav$ inferred from $T_{\rm eff}$ is approximately proportional to the
corresponding $L_{\rm q}$, fractional variations induced in $L_{\rm q}$ by a
time-variable $\dot M$ correspond directly to fractional uncertainty in an
inferred $\timav$.  Since Figure \ref{fig:var-time} shows the response to an
order-unity variation in $\dot M$, it also approximately quantifies the
fractional uncertainty for a given timescale of variation.  Assuming that
$\dot M$ has only short or moderate timescale variations, even up to $10^3$
yr, but is consistent on longer timescales, $T_{\rm eff}$ is a good indicator
of the time-averaged $\dot M$, with minimal uncertainty (20\%) at low $\dot
M$ and moderate (50\%) at high $\dot M$.  However, due to the long but finite
thermal time of the envelope, under the assumption that $\dot M$ varies over very
long timescales ($>10^3$ yr for high-$\dot M$ systems or $> 10^5$ yr for
low-$\dot M$ systems), $T_{\rm eff}$ is not a reliable indicator
of $\timav$, having a large uncertainty.  In this case $T_{\rm eff}$ depends
more strongly on the recent accretion history than the overall average.  Thus
the character of the mass-transfer law being tested must be considered when
$T_{\rm eff}$ measurements are used to infer $\timav$ values.  The utility of
this relationship between uncertainty and timescale and its application to
candidate mass-transfer scenarios is discussed in more detail in
\S\ref{sec:discussion}.  There, multiple $T_{\rm eff}$ measurements from
similar systems are used to obtain an independent constraint on the
time-variability of $\dot M$, removing the \emph{a priori} model dependence
for some systems.

%Since the heat release due to compression occurs on the local thermal time,
%a good approximation of the surface luminosity as a function of time is
%given by
%\begin{equation}
%L \simeq
%\int_0^{M_{\rm acc}} \timav(t,t_{\rm th}) \epsilon_{\rm comp} dM_r
%\end{equation}
%where
%\begin{equation}
%\timav(t,t_{\rm th}) = \frac{1}{t_{\rm th}}\int_{t-t_{\rm th}}^{t} \dot M\,dt
%\end{equation}
%and $\epsilon_{\rm comp}$ is the heat liberated by compression (TODO) and
%$t_{\rm th}$ is a function of $M_r$.

\section{System Features Affecting the Utility of $T_{\rm eff}$}
\label{sec:system}

In addition to long-timescale variations in the mass transfer rate,
several features of the systems in which the quiescent $T_{\rm eff}$
measurements can be made affect our ability to determine $L_{\rm q}$
from observations.  While we believe that these issues can be avoided
by careful selection and interpretation of observations, it is worth
summarizing the relevant issues to justify this assertion.
\citet{piroetal05-1} have performed an excellent analysis of how a
dwarf nova cools after outburst, finding that the cooling time depends
on the amount of matter accreted in the outburst.  We therefore forego
a detailed discussion of those systems, only noting that care is
typically taken that $T_{\rm eff}$ is measured as far in quiescence as
is feasible.  The other important kinds of systems in which $T_{\rm
eff}$ can be measured appropriately are novalike variables
during extended quiescence intervals and Magnetic CVs, again in quiescence,
where material impacts on a small portion of the WD surface, and the emission
from the rest of the star, reflecting $L_{\rm q}$, can be separated.  After
discussing these two cases we provide a direct estimate of the timescale on
which a CVWD would cool to $T_{\rm eff,q}$ after a thermonuclear runaway, and
close with a brief discussion of additional energy sources related to
accretion which could cause the $T_{\rm eff}$ observed to differ from our
calculated $L_{\rm q}$.

\subsection{Cooling From High State}

\begin{figure}
%\plotone{TTAri_examples.eps}
\plotone{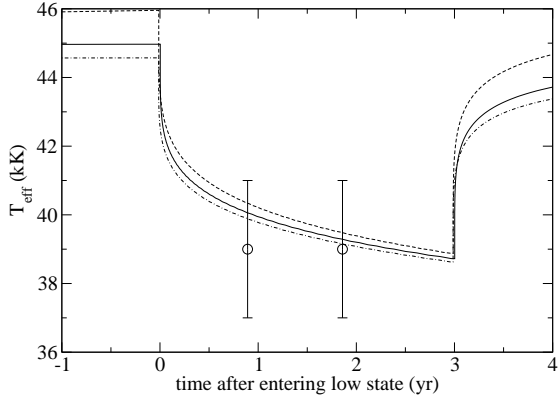}
\caption{
\label{fig:TTAri}
Decline of $T_{\rm eff}$ for the 3-year low state of the novalike TT Ari
during 1982-4.  UV spectral observations taken at approximately 1 and 2 years
after the decline into the low state are both consistent with $T_{\rm eff} =
39000\pm2000$ K \citep{gaensickeetal99-1}.  The three example cases shown
have the same $\timav=4.1\times 10^{-9}M_\odot$~yr$^{-1}$ but different high
state durations: 30 yr (\textit{solid}), 10 yr (\textit{dashed}), and 90 yr
(\textit{dot-dashed}).
}
\end{figure}

In the VY Scl stars discussed in section \ref{sec:vyscl}, the high
$\dot M\sim 10^{-8}M_\odot$~yr$^{-1}$ will, on rare occasions,
turn off for timescales of up to a few years.  This provides a window
in which to measure the $T_{\rm eff}$.  As discussed above in section
\ref{sec:timescales}, the layer which is heated by infall cools in
$\sim$100 seconds at this high $\dot M$ and
resulting high $T_{\rm eff,q}$.  Since the outer layers, which can change
their thermal state during quiescence, contribute a relatively small
fraction of the overall $L_{\rm q}$, the cooling during the quiescence
is modest.  In Figure~\ref{fig:TTAri} we have calculated an example
evolution for the $\sim$3-year quiescence of TT Ari during which it
was observed twice \citep{gaensickeetal99-1}, assuming various high
state durations between regularly repeated quiescent intervals.  With
this type of analysis we are able to solve directly for the actual
$\timav = 4.1\pm0.8\times 10^{-9}M_\odot$~yr$^{-1}$ for $M=0.9M_\odot$,
accounting for the
cooling of the outer layers during quiescence.  It is observed from
Figure~\ref{fig:TTAri} that the $T_{\rm eff}$ flattens out as the
thermal time of the cooling layer becomes longer, in such a way that
the two epochs of measurements (shown) will be the same within the
observational error.  By varying the high state interval at fixed
$\timav$, we see that, even for durations as short as 10 years, the
quiescent flux is a good indicator of $\timav$ instead of the
high-state $\dot M$.  The $L_{\rm q}$ scales linearly with $\timav$ as
expected.

\subsection{Polar Accretion}

\label{sec:polar}
The geometry of the accretion on the WD surface differs significantly
for a strongly magnetic WD ($B\sim 10^7$-$10^8$ G), in which case mass
is deposited at the magnetic poles. However, due to the depth at which
this material must spread over the star and the depths at which energy is
liberated by compression, the effect on the $T_{\rm eff}$ measured away from
the polar regions is modest.  The magnetic field can only constrain the
accreted material to the polar regions down to a limited pressure depth,
$P_{\rm crit}$, below which lateral pressure gradients are strong enough to
force the material to spread over the surface.  Heat liberated by compression
up to $P_{\rm crit}$ is localized to the polar regions, while that from
compression at higher pressures is spread over the whole surface.

The magnetic field can keep the accreted
material constrained to the WD poles up to a depth where $\beta_{\rm
crit} \simeq 2\ell/h$ \citep{Hameetal83} where $\beta = P/4\pi B^2$ is
the ratio of gas to magnetic field pressure, $\ell$ is the size of the
polar cap, and $h$ is its thickness.  Using $h= P/\rho g$,
$\ell=10^{8}$ cm and a degenerate equation of state we find that
$P_{\rm crit}\sim 10^{15}\ {\rm erg}\ {\rm cm^{-3}}\ (g_8 \ell_8
B_7^2)^{5/7}$, where a subscript $s$ indicate values divided by $10^s$
cgs units, e.g. $g_8 = g/10^8\ {\rm cm}\ {\rm s^{2}}$.  Note that a
nondegenerate calculation, which is more appropriate since degeneracy sets in
at $P\simeq 10^{17}$ erg cm$^{-3}$ for $T_c=10^7$ K, would give even lower
$P_{\rm crit}$.  We forgo such a calculation here since the difference does
not impact our conclusions.

\begin{figure}
%\plotone{Lr-P.eps}
\plotone{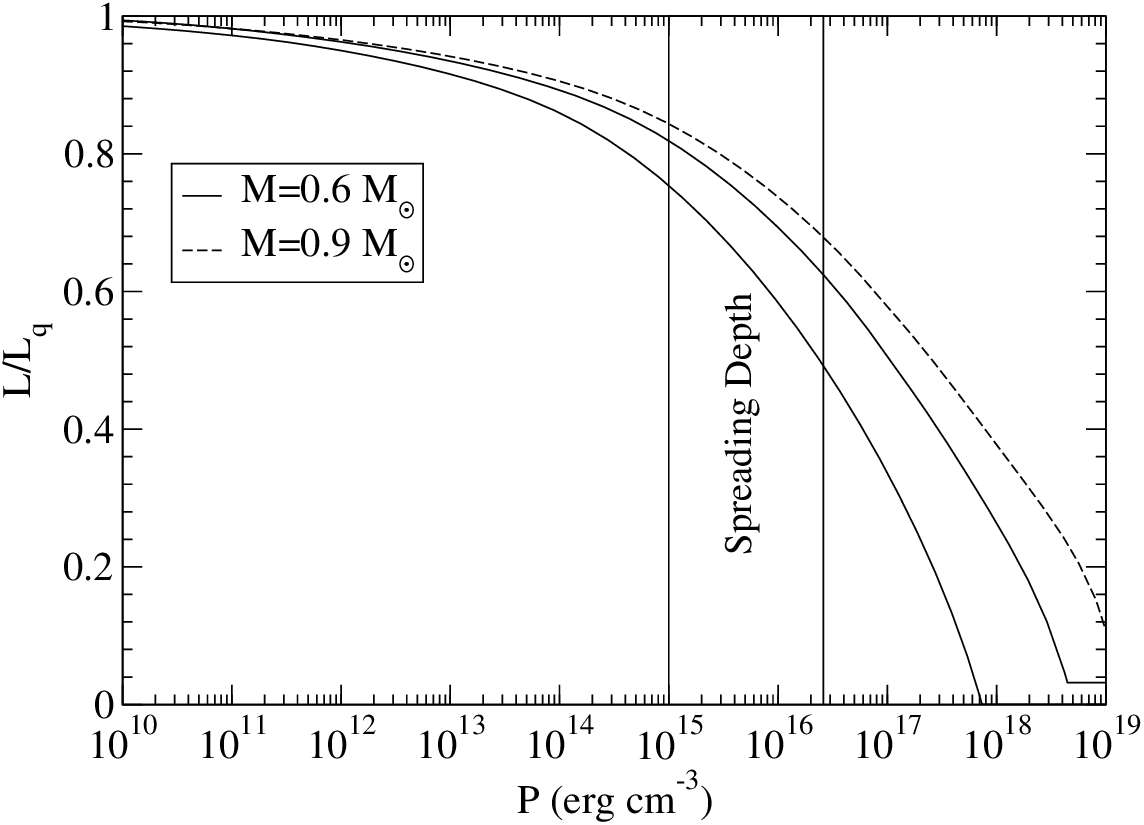}
\caption{\label{fig:Lr-P} Luminosity as a function of depth as
measured by $P$ as a fraction of $L_{\rm surf}$.  The three cases
shown are for $\timav =2\times 10^{-9}$ (lower) and $5\times 10^{-11}
M_\odot$ yr$^{-1}$ (upper) at $M=0.6M_\odot$ (solid) and only the
lower $\timav$ at $M=0.9M_\odot$ (dashed).  These have $L_{\rm q}$
($T_{\rm eff}$) of 140, 2.6 and 2.8 $\times 10^{-3}L_\odot$ (32, 12,
and 14 kK).  The bulk of the compressional luminosity is released below the
depth at which spreading occurs.}
\end{figure}

As discussed above, energy is liberated by compression at all depths, with
much coming from deep in the accreted layer.  Figure \ref{fig:Lr-P} shows how
$L/L_{\rm q}$ varies with depth into the star as measured by $P$.  In a
magnetic CV WD, we assume that accreted matter spreads over the WD surface at
approximately $P_{\rm crit}$, so that only $L|_{P=P_{\rm crit}}$ is released
over the whole surface of the star.  For typical fields of $10^7$ G, this
amounts to better than 80\% of the $L_{\rm q}$ expected in the non-magnetic
case.  Thus magnetic CV $T_{\rm
eff}$'s warrant a small upward correction from the simple relation given by
eq.~(\ref{eq:Teffrelation}) to obtain the actual $\timav$.

\subsection{Cooling After Thermonuclear Runaway}
\label{sec:aftercn}

After shutoff of the nuclear burning in a classical nova, due to heat
leftover from the outburst it will take some period of time for $T_{\rm eff}$
to return to that characteristic of $\timav$.  Although this timescale could
vary widely depending on how much of the burning envelope is actually left
behind on the WD, it appears that in most cases the ejected matter is similar
to or larger than the total accreted, indicating that this layer is fairly
thin (TB04).  The enhancement of the temperature below the burning layer is
expected to be modest, being limited to just above $10^7$ K by the onset of
the instability.  A useful estimate for the cooling time is then to take the
cooling time of the outer radiative layer.  Using $L_{\rm q}$, the endpoint
luminosity of the decline, to estimate this cooling time gives a good upper
limit on the decline from the hot outburst state.

The mass coordinate, measured in from
the surface, where the free-free opacity and that due to conduction are equal
is $\Delta M \simeq 4\times 10^{-6}M_\odot (M/0.9M_\odot)^{-8.2}$ for an
interior temperature of $10^7$ K.  This leads to a thermal time of the
radiative layer of
\begin{equation}
\tau_{\rm th,rad}
\simeq 7\times 10^4{\rm\ yr}\left(\frac{T_{\rm eff}}{14{\rm kK}}\right)^{-4}
\left(\frac{M}{0.9M_\odot}\right)^{-4.6}\ ,
\end{equation}
where we have used the approximate dependence $R\propto M^{-1.8}$.  Note that
this is similar to the timescale at which the response curves from above the
variation analysis (see Figure \ref{fig:var-time}) flatten off.
This is reasonably consistent with the timescale found by \citet{prialnik86-1}
from nova simulations at high mass, $M=1.25M_\odot$.
This is generally much shorter than the inter-outburst time and therefore we
expect few objects to show such an enhancement.
If the heated region extends deep enough for some degenerate material to have
a longer cooling time, which may be the case at low $\timav$, it would
introduce an enhanced lower boundary temperature for the fresh envelope for a
longer time.  However, this will not lead to significant enhancement of
$L_{\rm q}$ because $L_{\rm q}$ is much larger than the cooling luminosity
set only by the modestly increased temperature at the base of the layer.

\subsection{Possible Additional Heat Sources}
\label{sec:additional}

If there is a heat source depositing excess energy at the WD surface
during quiescence, this can lead to misinterpretation of the quiescent
$T_{\rm eff}$ as representative of $L_{\rm q}$.  The energy release due to
infall is much larger than that due to compression in the WD, so that
a small amount of steady accretion during quiescence in non-magnetic
systems could compete with the flux from below in heating the WD surface.
Most of the energy from quiescent accretion is expected to
appear in the X-rays and typical X-ray luminosities from dwarf novae
are $10^{30}-10^{31}$~erg~s$^{-1}$ \citep{verbuntetal97-1} requiring a
quiescent $\dot M\sim 10^{-12}M_\odot$~yr$^{-1}$.  While this is
clearly less than the luminosity of the brighter systems we will
consider, it is only slightly less than the lowest $L_{\rm q}\sim
10^{31}$ erg s$^{-1}$ ($T_{\rm eff} =14$ kK, $M=0.9M_\odot$) that we are
considering here.  We will ignore this contribution in the first
approximation for two reasons.  First, it is not expected that $L_{\rm
UV}$ and $L_{\rm X}$ should be similar, if anything $L_{\rm UV}$
should be much lower, if there is a steady state corona above the WD
surface where the infall energy is released.  Any heating which
modifies $T_{\rm eff}$ must be deposited \emph{at the photosphere},
and no higher.  Second, while there is a wide variety of X-ray
luminosities for dim DN which has no clear cause, we will show that
the $T_{\rm eff}$ for these systems show relatively little
scatter.  At worst our dimmest systems provide upper limits on the
$\timav$ if some portion of the flux is attributed to accretion in
quiescence.

\section{Measurements of CV Primary $T_{\rm eff}$}
\label{sec:measurements}

The most commonly used method to determine WD effective temperatures
in CVs is to fit synthetic spectra to optical and/or ultraviolet (UV)
observations. While for single white dwarfs accurate temperatures and
good estimates of the surface gravity (and hence, adopting a
mass-radius relationship, the WD mass) are routinely derived from
spectral fits to the Balmer lines alone, a number of caveats have to
be considered when applying such spectral fits to the observations of
CVWDs. One main disadvantage encountered in CVs is that their optical
light is a mixture of emission from the accretion disk or stream, hot
spots on the disk edge or WD surface, from the donor star, and finally
from the accreting WD. In the majority of CVs, the WD contributes only
a small fraction to the emission at optical wavelengths, which foibles
any attempt to determine its properties from ground-based
observations. Even in the cases where the WD is a significant source
of optical flux, implying low prevailing accretion rates, temperature
measurements are subject to a major ambiguity: the strength of the
Balmer lines reaches a maximum around 15\,000\,K, with the exact value
being a function of surface gravity, and nearly equally good fits can
be achieved on the "hot" and the "cold" side. 

Because the WDs in CVs are moderately hot, $10\,000-50\,000$\,K, their
spectral energy distribution peaks in the UV, and therefore
the most reliable information on CVWDs is obtained from
space-based telescopes such as the \textit{IUE}, \textit{HST}, or
\textit{FUSE}. Besides a larger, or dominant, WD contribution to the
total UV flux from the CV compared to the optical, the
degeneracy between hot and cold Balmer line fits is broken by the
opacity of quasimolecular H$_2^+$ and H$_2$, which causes broad
absorption lines near 1400\,\AA\ and 1600\,\AA\ for temperatures below
$\sim18\,000$\,K and $\sim13\,000$\,K, respectively
\citep{koesteretal85-1}.  Examples of the degree of uncertainty in
temperatures based on optical data are GW\,Lib and LL\,And, where
\citet{Szkoetal00} and \citet{szkodyetal00-3} estimated
$\Teff\simeq11\,000$\,K from optical data obtained during quiescence,
whereas the temperatures determined from \textit{HST}/STIS
spectroscopy are $14\,000-15\,000$\,K
\citep{szkodyetal02-4,howelletal02-1}. 
While UV spectroscopy can provide fairly accurate WD effective
temperatures, its diagnostic potential for determining the surface
gravity, $\log g$, is very limited, and so is, therefore, the ability
to derive the white dwarf mass from spectral fits.

A remaining issue
in modelling the UV data of accreting white dwarfs is that a large
fraction display a second continuum flux component, that typically
contributes $10-30$\%. A number of suggestions for the nature of this
component have been made, such as optically thick hot accretion belts
on the WD (e.g. \citealt{longetal93-1,
gaensicke+beuermann96-1,huangetal96-1}), optically thick emission from
the hot spot or optically thin emission from a chromosphere on the
accretion disk (e.g. \citealt{gaensickeetal05-2}). While the exact
nature of this additional component is not clear, and may well differ
among the objects, its impact on the effective temperature
determination appears to be only modest. 

An alternative method that yields WD effective temperatures, as well
as potentially radii and masses as well, is the modelling of the WD
ingress/egress in multi-color light curves of eclipsing CVs
\citep{wood+horne90-1}. The application of that method has been
limited until recently to a handful of bright CVs, but since the fast
triple-beam CCD camera ULTRACAM \citep{dhillonetal07-1} became
available on 4\,m and 8\,m telescopes, a number of detailed CVWD
studies have been carried out (e.g. \citealt{littlefairetal06-2}).

The literature is humming with values of CVWD temperatures and some
care has to be taken to differentiate between measurements and
estimates. For some purposes, it may be desirable to maximize the
number of available $\Teff$ values, such as provided e.g. by
\citet{WintSion03} and \citet{urban+sion06-1}. In the context of providing
a stringent test of the theory angular momentum loss in CVs, we focus
here on the most accurate $\Teff$ measurements. Table\,\ref{t-teff}
lists CVWD $\Teff$ values that we consider reliable on the basis that
the WD has unambiguously been detected either spectroscopically, or,
in the case of eclipsing systems, through its eclipse ingress/egress.
Below in Sect.\,\ref{sec:dn}--\ref{sec:am}, we discuss particular
issues that relate to the effective temperature measurements in the
three major CV subclasses, and summarize in Sect.\,\ref{sec:failed}

\subsection{Dwarf novae}
\label{sec:dn}
Dwarf novae are a subset of non-magnetic CVs with low mass transfer
rates. The accretion disks in these systems are thermally unstable,
and undergo outbursts lasting a few days to a few months with
recurrence times of a few weeks to tens of years. Despite their
relatively low mass transfer rates, only a relatively small number of
dwarf novae reveal their accreting white dwarfs at optical
wavelengths, e.g. WZ\,Sge \citep{greenstein57-1} or VW\,Hyi
\citep{mateo+szkody84-1,smithetal06-1}. Moving to the UV, about
a third of all short-period dwarf novae are dominated by emission from
the white dwarf. It is currently unclear why dwarf novae with nearly
identical orbital periods and outburst frequencies differ radically in
the characteristics of their UV spectra, such as e.g. VW\,Hyi and
WX\,Hyi, with the first one being one of the best-studied CVWDs
\citep{mateo+szkody84-1, sionetal95-3, gaensicke+beuermann96-3,
longetal96-2, smithetal06-1} and WX\,Hyi, where no convincing
spectroscopic signature from the WD has been detected
\citep{longetal05-1}. The fraction of dwarf novae above the period gap
where the WD is clearly discernible in the UV is much smaller than at
short orbital periods, as the accretion disks in these systems are
larger and can sustain higher accretion rates while still remaining in
quiescence.

During dwarf nova outbursts, the accretion rate onto the white dwarf
increases by several orders of magnitude with respect to the quiescent
value, causing a short-term heating effect \citep{Sion95,
gaensicke+beuermann96-1, chengetal00-1, godonetal04-1, piroetal05-1}. Therefore,
when determining the secular WD effective temperature, care has to
be taken to observe dwarf novae as long as possible after an
outburst. In systems with very short outburst recurrence times, it may
be that the WD never cools to its secular
temperature \citep{gaensicke+beuermann96-1}.  (See also the discussion in
\S\ref{sec:reachingquiescent}.)

A final caveat relates to the spectral modelling of high-inclination
dwarf novae, where the line-of-sight passes through absorbing material
located above the accretion disk, termed an accretion veil, affecting the
effective temperature determination. In the case of OY\,Car, a fit ignoring
the veiling component yields $T_\mathrm{eff}\approx15\,000$\,K
\citep{horneetal94-1}, whereas $T_\mathrm{eff}\approx17\,000$\,K when
taking the absorption into account \citep{horneetal94-1,
chengetal00-1}. While the effect is noticeable at inclinations
$\ga70^\circ$ (e.g. \cite{longetal06-1}, it is most problematic at
higher inclinations where the WD eclipse offers an
alternative/independent possibility of a $T_\mathrm{eff}$
measurement (e.g. \citealt{littlefairetal06-1}).

\subsection{Novalikes and the case of VY\,Scl stars}
\label{sec:vyscl}
Novalike variables are non-magnetic CVs, mainly with periods $>3$\,h,
with high mass transfer rates in which the accretion disk is in a
stable hot and optically thick state. Consequently, the flux of
novalike variables is entirely dominated by the disk at optical and
ultraviolet wavelengths. The only possibility to learn about the
properties of WDs in novalike variables occurs if the mass transfer
decreases, or turns off completely, so that the WD becomes
visible. Novalike variables who show such \textit{low states} are
called VY\,Scl stars, after the prototypical system, and are found
predominantly in the orbital period range $3-4$\,h. The physical cause
of the occurrence of low states is not fully understood, and may be
related to star spots on the secondary star \citep{livio+pringle94-1,
hessmanetal00-1} or irradiation driven mass transfer cycles
\citep{wuetal95-3}. 

Due to the rare and unpredictable nature of low states, only three
novalike variables have been studied at a sufficient level of detail:
TT\,Ari \citep{shafteretal85-1, gaensickeetal99-1}, DW\,UMa
\citep{kniggeetal00-1, araujo-betancoretal03-1} and MV\,Lyr
\citep{hoardetal04-1}. In all three systems, hot
$T_\mathrm{wd}\ga40\,000$\,K are found. 

\subsection{Polars}
\label{sec:am}
Polars, or AM\,Herculis stars, contain strongly magnetic WDs. The
rotation of the white dwarf is synchronized with the orbital period,
and the formation of an accretion disk is suppressed. Accretion occurs
via an accretion stream that feeds matter onto the magnetic pole of
the WD. Polars enter low states with little or no accretion, and
during these episodes the systems appear practically as a detached WD
plus main sequence star. However, while the white dwarf is fully
exposed during low states, its magnetic field complicates accurate
$T_\mathrm{eff}$ determinations, as the Balmer lines are subject to
Zeeman splitting, and no accurate theory for the line profiles of the
Zeeman components exists so far \citep{jordan92-1}. Zeeman splitting
is much weaker for the Lyman lines, and for fields $\la30$\,MG the
effect of the magnetic field on temperatures obtained from UV
observations around Ly$\alpha$ is relatively small. The polars in
Table\,\ref{t-teff} have all fields $\la30$\,MG at the primary
accretion pole, with the exception of QS\,Tel and V1043\,Cen (both
$B\simeq56$\,MG).

For stronger fields, even the Lyman lines become useless for
temperature estimates, and, worse, the spectral models fail to even
reproduce the UV/optical spectral energy distribution, an effect known
from single white dwarfs \citep{schmidtetal86-2}. Consequently, the WD
temperatures of high field polars are only very approximatively known,
$15\,000-25\,000$\,K for AR\,UMa  \citep{GansSchm01} and
$17\,000-23\,000$\,K for RX\,J1554.2+2721
\citep{gaensickeetal04-3}. 

The highly asymmetric accretion geometry in polars results in heating
part of the WD atmosphere around the magnetic pole(s)
\citep{gaensickeetal95-1}. As the magnetic axis of the WD is usually
not aligned with its spin axis, the heated pole cap acts as a light
house, causing a significant variation of the UV flux as a function of
WD spin/orbital phase. The pole caps are observed also during low
states, and it is not clear if this is due to deep heating, or to
residual low-level accretion in the low state \citep{stockmanetal94-1,
gaensickeetal95-1}. In some polars, the geometry is favorable and the
heated pole cap is eclipsed by the body of the white dwarf for part of
the orbital cycle, allowing an accurate determination of
$T_\mathrm{eff}$ from phase-resolved spectroscopy
\citep{gaensickeetal06-2}. If the pole cap contributes at all phases
to the UV light, or only phase-averaged data is available, the data
can be fit with a two-component model
\citep[e.g.][]{gaensickeetal00-1,
  araujo-betancoretal05-2}. $T_\mathrm{eff}$ from such analyses
provides an upper limit to the true WD temperature. 

\subsection{Reliable $T_{\rm eff}$ measurements}
\label{sec:failed}
In the context of using $\Teff$ as a measurable quantity that allows
insight into the secular averages of the mass transfer rates in CVs,
we have included in Table\,\ref{t-teff} only those systems which we
feel have a reliable $\Teff$ determination. Consequently, we omitted
systems with published WD temperatures where the evidence for seeing
the WD is ambiguous. Examples of such cases are WX\,Hyi, SS\,Cyg, and
RU\,Peg \citep{sion+urban02-1, longetal05-1}, where a plausible WD
model fit to the UV spectra can be achieved, but no clear WD features
are discerned (broad Lyman lines, narrow metal absorption lines). The
decision of admitting a system to Table\,\ref{t-teff} is necessarily
subject to a gray area, where some spectroscopic evidence for the WD
is present, but not sufficient for an accurate $T_\mathrm{eff}$
determination, such as e.g. the case of Z\,Cam
\citep{hartleyetal05-1}. \citet{gaensicke+koester99-1} have shown that
in the case of low spectral resolution and low signal-to-noise the UV
data may be equally well described by a moderately hot WD or by an
optically thick accretion disk. In this particular case, AH\,Men, the WD case can be
excluded on the basis of the optical properties of the star, providing
a clear warning against interpreting a slight flux turnover below
1300\,\AA\ as broad L$\alpha$ from a WD photosphere.

In the case of eclipse light curve analyses, we excluded a number of
systems where we considered the data of too low quality, i.e. an at
best marginal detection of the WD ingress/egress, as well as studies
using oversimplified models, such as approximating the WD emission in
the different observed wave bands by blackbody radiation.

In recent years, a number of strongly magnetic close WD+MS binaries
were identified that contain very cool ($\Teff\la9000$\,K) WDs and
have mass transfer rates, as determined from X-ray observations, of a
few $10^{-13}\dot M\,\mathrm{yr^{-1}}$
\citep[e.g.][]{reimers+hagen00-1, szkodyetal03-3, schmidtetal05-1,
vogeletal07-1}.  Given the fact that many of them have MS companions
that have spectral types too late to be Roche-lobe filling at
the orbital periods of the binaries,
\citet{webbink+wickramasinghe05-1} suggested that these systems are
pre-CVs that have not yet evolved into a semi-detached
configuration. The low mass transfer rates are compatible with wind
accretion, and the low WD temperatures match with the predictions for
the average life time of pre-CVs
\citet{schreiber+gaensicke03-1}. Consequently, we exclude those
systems from the present discussion.

A final note concerns the intermediate polars (IPs), a class of weakly
magnetic CVs in which the WD spin period is shorter than the orbital
period, and partial accretion disks may form. In most IPs, the
accretion rate is too high to discern the WD even at UV wavelengths
\citep[e.g.][]{mouchetetal91-1, beuermannetal04-1}. In a few IPs,
moderately broad Balmer absorption lines were detected in the optical
\citep[e.g.][]{haberletal02-1} but a more detailed analysis of the
system parameters, in particular the distance, ruled out a WD
photospheric origin of these features \citep{demartinoetal06-1}
In the case of EX\,Hya \citep{eisenbartetal02-1, belleetal03-1} and
AE\,Aqr \citep{eracleousetal94-1} \textit{HST} spectroscopy provided
more convincing evidence for the detection of thermal emission from
the WD photosphere, compatible with temperatures $\sim25\,000$\,K,
however, at least in the case of AE\,Aqr that temperature was cleary
not that of the quiescent WD, but of the accretion-heated pole-cap. In
summary, we did not include any IP in Table\,\ref{t-teff} because of
the lack of a clear detection of the quiescent WD in any of these
objects. 

\begin{table*}
\caption{\label{t-teff}Reliable $T_\mathrm{eff}$ measurements for WDs
  in CVs. Also listed are the CV subtype, the orbital period,
  the distance implied by the WD fit, the WD mass if measured
  independently from the spectral fit, and the distance obtained from
  a trigonometric parallax.}
\newcounter{ref}
\newcommand{\tcite}{\stepcounter{ref}\arabic{ref}}
\newcommand{\tref}[1]{\stepcounter{ref}(\arabic{ref})\,\citealt{#1}}

\begin{center}
\begin{tabular}{llrrrlrrl}
\hline\hline\noalign{\smallskip}
System & Type  & \Porb\,[h] & 
\Teff\,[K] & $\pm$  & \multicolumn{1}{c}{d\,[pc]} &
\multicolumn{1}{c}{\Mwd\,[\Msun]} & \multicolumn{1}{c}{d\,[pc]} & Ref \\
\noalign{\smallskip}\hline\noalign{\smallskip}
GW\,Lib     & DN/WZ & 1.280 & 14700 &      & 150--170    &             & $104\pm{30\atop20}$   & \tcite,\tcite \\ 
BW\,Scl     & DN\,? & 1.304 & 14800 &  900 & $131\pm18$  &             &                       & \tcite \\
LL\,And     & DN/WZ & 1.321 & 14300 & 1000 & $760\pm100$ &             &                       & \tcite \\
EF\,Eri     & AM    & 1.350 &  9500 &  500 & $\simeq130$ &             & $163\pm{66\atop50}$   & \tcite,\tcite, 2 \\
SDSS\,J1610-0102 & DN? & 1.34 & 14500 & 1500 &           &             &                       & \tcite \\
HS2331+3905 & DN & 1.351 & 11500 &  750 &  $95\pm15$  &             &                       & \tcite \\
AL\,Com     & DN/WZ & 1.361 & 16300 & 1000 & $800\pm150$ &             &                       & \tcite \\ 
WZ\,Sge     & DN/WZ & 1.361 & 14900 &  250 & 69          & $0.85\pm0.04$ & $44\pm2$              & \tcite,\tcite,\tcite,\tcite,\tcite, 2 \\
SW\,UMa     & DN/SU & 1.364 & 13900 & 900  & $159\pm22$  &             &                       & 3 \\
SDSS\,J1035+0555 & DN? & 1.368 & 10500 & 1000 &          & $0.94\pm0.01$ &                     & \tcite,\tcite  \\
HV\,Vir     & DN/WZ & 1.370 & 13300 & 800  & $480\pm70$  &             & $460\pm{530\atop180}$ & \tcite, 2 \\  
WX\,Cet     & DN/WZ & 1.399 & 13500 &      & 133         &             &                       & \tcite \\  
EG\,Cnc     & DN/WZ & 1.410 & 12300 & 700  & $420\pm60$  &             &                       & 15, 2 \\
XZ\,Eri     & DN/SU & 1.468 & 15000 & 1500 &             & $0.767\pm0.018$&                      & \tcite \\
DP\,Leo     & AM    & 1.497 & 13500 &      & 400         & $\sim0.6$   &                       & \tcite\\
V347\,Pav   & AM    & 1.501 & 11800 & 600  & $177\pm{33\atop38}$ &     &                       & \tcite \\  
BC\,UMa     & DN/SU & 1.503 & 15200 & 1000 & $285\pm42$  &             &                       & 3 \\  
EK\,TrA     & DN/SU & 1.509 & 18000 & 1200 & 200         &             &                       & \tcite \\ 
VY\,Aqr     & DN/WZ & 1.514 & 14500 &      & 187         &             & $97\pm{15\atop12}$    & 18, 2 \\  
OY\,Car     & DN/SU & 1.515 & 15000 & 2000 & $90\pm5$    &             &                       & \tcite,\tcite,\tcite \\ 
%UV\,Per     & DN/SU & 1.56  & 14200 & 1000 &             &             &                       & \tcite \\
%CY\,UMa     & DN/SU & 1.631 & 14700 & 1000 &             &             &                       & 19\\
%RZ\,Sge     & DN/SU & 1.639 & 14900 & 1000 &             &             &                       & 19\\
VV\,Pup     & AM    & 1.674 & 11900 & 600  & $151\pm{23\atop34}$ &     &                       & 21 \\
V834\,Cen   & AM    & 1.692 & 14300 & 900  & $144\pm{18\atop23}$ &     &                       & 21 \\
HT\,Cas     & DN/SU & 1.768 & 14000 & 1000 &             &             &                       & \tcite,\tcite,\tcite\\
VW\,Hyi     & DN/SU & 1.783 & 20000 & 1000 &             & $0.71\pm{0.18\atop0.26}$ &          & \tcite,\tcite,\tcite \\
%V844\,Her   & DN/SU & 1.884 & 14500 & 1000 &             &             &                       & 19 \\ 
CU\,Vel     & DN/SU & 1.88  & 18500 & 1500 &             &             &                       & \tcite \\
MR\,Ser     & AM    & 1.891 & 14200 &  900 & $160\pm{18\atop26}$ &     &                       & 21 \\
BL\,Hyi     & AM    & 1.894 & 13300 &  900 & $163\pm{18\atop26}$ &     &                       & 21 \\
ST\,LMi     & AM    & 1.898 & 10800 &  500 & $115\pm22$  &             &                       & 21 \\
EF\,Peg     & DN/WZ & 2.00: & 16600 & 1000 & $380\pm60$  &             &                       & 4 \\  
DV\,UMa     & DN/SU & 2.138 & 20000 & 1500 &             & $1.041\pm0.024$ &                   & 19 \\
HU\,Aqr     & AM    & 2.084 & 14000 &      &             &             &                       & \tcite\\
QS\,Tel     & AM    & 2.332 & 17500 & 1500 &             &             &                       & \tcite \\
SDSS\,J1702+3229 & DN/SU & 2.402 & 17000 & 500 & $440\pm30$ & $0.94\pm0.01$ &                  & \tcite \\
%TU\,Men     & DN/SU & 2.813 & 28000 &      & 210         &             &                       & \tcite \\
AM\,Her     & AM    & 3.094 & 19800 & 700  & $90$        &             & $79\pm{8\atop9}$      & \tcite,\tcite,2\\
MV\,Lyr     & NL/VY & 3.176 & 47000 &      & $505\pm50$  &             &                       & \tcite\\
DW\,UMa     & NL/VY & 3.279 & 50000 & 1000 & $590\pm100$ & $0.77\pm0.07$ &                     & \tcite\\
TT\,Ari     & NL/VY & 3.301 & 39000 &      & $335\pm50$  &             &                       & \tcite\\
V1043\,Cen  & AM    & 4.190 & 15000 &      & 200         &             &                       & \tcite\\
WW\,Cet     & DN    & 4.220 & 26000 & 1000 &             &             &                       & \tcite\\
U\,Gem      & DN/UG & 4.246 & 30000 & 1000 &             & $\sim1.1$   & $104\pm4$             & \tcite,\tcite,\tcite,13\\
SS\,Aur     & DN/UG & 4.391 & 27000 &      &             &             &                       & \tcite\\
V895\,Cen   & AM    & 4.765 & 14000 & 900  & $511\pm{60\atop81}$ &     &                       & 21 \\
RX\,And     & DN/ZC & 5.037 & 34000 & 1000 &             &             &                       & \tcite\\
%RU\,Peg     & DN/UG & 8.990 & 53000 &      &             &             &                       & 31\\
%
\noalign{\smallskip}
\hline		
\end{tabular}	
\end{center}
%\linebreak		
\setcounter{ref}{0} 
\tref{szkodyetal02-4}, 
\tref{thorstensen03-1},
\tref{gaensickeetal05-2}, 
\tref{howelletal02-1}, 
\tref{szkodyetal06-2},
\tref{beuermannetal00-1},
\tref{szkodyetal07},
\tref{araujo-betancoretal05-1},
\tref{szkodyetal03-1},
\tref{sionetal95-2},
\tref{steeghsetal01-2},
\tref{longetal04-1},
\tref{harrisonetal04-3},
\tref{steeghsetal07-1},
\tref{southworthetal06-1},
\tref{littlefairetal06-2},
\tref{szkodyetal02-3},
\tref{sionetal03-1},
\tref{felineetal04-1},
\tref{schwopeetal02-1},
\tref{araujo-betancoretal05-2},
\tref{gaensickeetal01-3},
\tref{hessmanetal89-1},
\tref{horneetal94-1},
\tref{chengetal00-1},
\tref{woodetal92-1},
\tref{woodetal95-2},
\tref{felineetal05-1},
\tref{gaensicke+beuermann96-1},
\tref{sionetal96-1},
\tref{smithetal06-1},
\tref{gaensicke+koester99-1},
\tref{gaensicke99-1},
\tref{rosenetal01-1},
\tref{littlefairetal06-1},
\tref{gaensickeetal95-1},
\tref{gaensickeetal06-2},
\tref{hoardetal04-1},
\tref{araujo-betancoretal03-1},
\tref{gaensickeetal99-1},
\tref{gaensickeetal00-1},
\tref{godonetal06-1},
\tref{long+gilliland99-1},
\tref{longetal06-1},
\tref{sionetal98-1},
\tref{sionetal04-1},
\tref{sionetal01-1}
\end{table*}

\begin{figure*}
%\plotone{Teff-Porb_wide.eps}
%\includegraphics[width=\textwidth]{Teff-Porb_wide.eps}
\includegraphics[width=\textwidth]{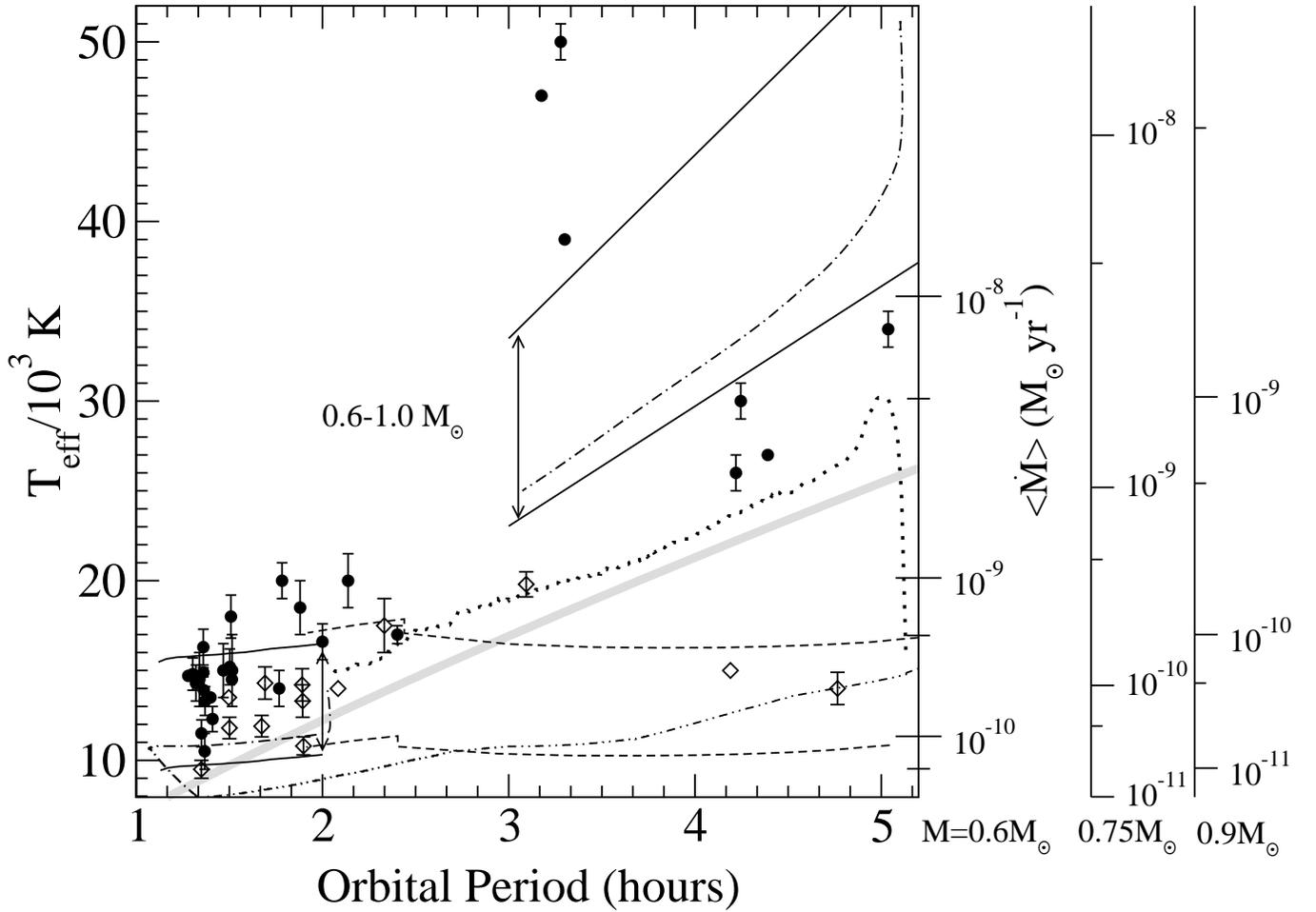}
\caption{
\label{fig:Teff-Porb_wide}
Observed $T_{\rm eff}$ in the best observational cases (data from Table\,
\ref{t-teff}).  Non-magnetic systems, a mix of dwarf novae and novalikes in
low states, are indicated with solid circles and magnetic systems, all
Polars, with open diamonds.  Error bars indicate uncertainty dominated by the
unknown WD mass.
Points without error bars are
\emph{less} accurate than those with, but their uncertainty is not
easily quantified.
An approxmate mapping to $\timav$ is shown on the right vertical scale
assuming $M=0.75M_\odot$, $0.6M_\odot$ or $0.9M_\odot$ as indicated.
The empirical relation of Patterson (1984; thick grey line)
is shown, along with several theoretical predictions: traditional magnetic
braking (\citealt{Howeetal01}; between solid lines, dot-dashed line),
Andronov et al.\ (2003; dot-dot-dash line),
Ivanova \& Taam (2004; dotted line), and evolution under only gravitational
radiation losses (between dashed lines).
There is a clear demonstration
that, for $P_{\rm orb}> 3$ hours, stellar wind angular momentum loss is
inhibited in the strongly magnetic systems.
}
\end{figure*}

%%%%%%%%%%%%%%%%%%%%%%%%%%%%%%%%%%%%%%%%%%%%%%%%%%%%%%%%%%%%%%%%55
\section{Discussion}
\label{sec:discussion}

While we will now inspect the values listed in Table\,\ref{t-teff} for
possible correlations, we must be utterly aware of the fact that the
set of known CVWD temperatures is subject to severe selection
effects. A very obvious, but crucial statement is that we need to be
able to see the WD in order to measure its temperature, and, as
mentioned above, this is not, or only marginally the case in systems
where the mass transfer rate is too high. Therefore, it is possible that
the temperatures obtained are rather lower limits than average values,
as WDs in systems with higher accretion rates will be hotter~--~but
not visible. This is most likely a stronger bias above the orbital
period gap than below.

Our collection of secure $T_{\rm eff}$ measurements as listed in Table
1 are plotted against system $P_{\rm orb}$ in
Figure~\ref{fig:Teff-Porb_wide}.  Non-magnetic (solid circles) and
strongly magnetic (open diamonds) systems are differentiated by symbol
type.  For the highest quality measurements, the uncertainty in
$T_{\rm eff}$, which is dominated by the unknown WD mass, is indicated
with an error bar.  Note that this means points without error bars are
\emph{less} accurate than those with, but their uncertainty is not
easily quantified.  An approximate $\timav$ scale is shown on the
right if an expected typical mass (0.75$M_\odot$;
\citealt{smith+dhillon98-1}; \citealt{knigge06-1}) is chosen, but the degree of scatter
expected is essentially unknown.
We show for comparison the empirical relation of \citet{patterson84-1}, using an
assumed primary mass of $M=0.75 M_\odot$.  From our $T_{\rm eff}$ data we
infer modestly higher mass transfer rates at all orbital periods, although a
higher average mass is also a viable explanation.

It is useful to dislay a variety of theoretical predictions to compare with
our data.
Solid lines show the $T_{\rm eff}$
range expected for the $\timav$ obtained from typical interrupted
magnetic braking scenarios, for the mass range 0.6 to
$0.9M_\odot$ (see \citealt{Howeetal01} and references therein).
Here we have used the mass-radius relation for the donor
star in the study by \citet{kolb+baraffe99-1} for below the period gap and
above the gap a value of $\timav = 10^{-9}M_\odot$ at $P_{\rm orb}=3$
hours and $10^{-8}M_\odot$~yr$^{-1}$ at $P_{\rm orb}=6$ hours, with a
line between.  Also shown is a curve (dot-dashed) for the $M=0.7M_\odot$
history from \citet{Howeetal01}.
The relation of \citet{Andretal03} for an unevolved donor is also shown, but
implies much lower $\timav$ than the data support.
The mass-transfer history of a case utilizing the weaker braking law
posited by \citet{IvanTaam04} is shown by the dotted line.  This curve
corresponds to their $\timav$ history for a $0.8M_\odot$ donor and a $0.6
M_\odot$ primary, but for display purposes we have used $M=0.75 M_\odot$ for
the primary mass to obtain the $T_{\rm eff}$, thus the true predicted $T_{\rm
eff}$ is likekly slightly higher.
For comparison to the magnetic systems,
we have used the main
sequence mass-radius relation given by \citet{Howeetal01} to determine
$\timav$ for non-conservative mass transfer under gravitational
radiation only.  The resulting $T_{\rm eff}$ is shown by dashed lines,
again for the mass range 0.6 to $0.9M_\odot$.

Uncertainty and caveats due to selection bias will be discussed below, as
will the contrast between magnetic and non-magnetic systems.  Comparing our
measurements for non-magnetic systems to the predictions shows that we infer
mass transfer rates that are lower than the predictions of traditional
magnetic braking.  Assuming that there is no strong selection bias toward
lower $\timav$ systems, not a trivial assumption as discussed below, our data
above the period gap are more consistent with the weakened angular momentum
loss proposed by \citet{IvanTaam04}, though still a bit higher than their
predictions.  The angular momentum law in the traditional magnetic braking
picture was largely calibrated in order to raise the radius of the companion
enough to reproduce the period gap, and, by virtue of this fact, evolution
following the \citet{IvanTaam04} relation might lack an appropriate period
gap.  The consistency of our data with the "softer"
law of \citet{IvanTaam04} and the presence of the high-$\timav$ VY Scl
systems, suggest a very different picture, where the bloating of the
companion might be localized to just above the period gap.
There remains much room for improvement in the development of braking laws,
as current laws remain fairly empirical with only modest input from
understanding of the magnetospheric structure and even less from possible
properties of the stellar dynamo.

\subsection{Uncertainty, Scatter, and Bias}

We would like to utilize our collection of $T_{\rm eff}$ measurements mainly
to constrain the dependence of $\timav$ on $P_{\rm orb}$ and thereby evaluate
angular momentum loss laws utilized to predict this relation in CV
evolution.  There are two important sources of general uncertainty bearing on
conclusions drawn from this dataset, which follow from the discussions of
previous sections: (1) uncertainty due to the effects of long-term $\dot M$
variations on $T_{\rm eff}$, and (2) uncertainty due to inaccessibility of
quiescent $T_{\rm eff}$ measurements at a given $P_{\rm orb}$.  Each of these
will introduce qualifications to the naive interpretation that $L_{\rm q}$ 
can be converted directly into $\timav$ representative of that $P_{\rm orb}$
interval for that kind of system (magnetic or non-magnetic) via a relation
like equation \ref{eq:lumrelation}.

\subsubsection{Long-term $\dot M$ Variations}
\label{sec:longmdotvary}

Given only a single $T_{\rm eff}$ measurement, it is always possible that
$\timav$ is something other than that implied by $T_{\rm eff}$, and we are
observing a transient state.  However, we do not have only one $T_{\rm eff}$
measurement; we have multiple measurements for objects in each of the 3
classes discussed in section \ref{sec:measurements}: dwarf novae, novalikes
displaying low states, and polars.  In light of the time series and analysis
of section \ref{sec:theory}, medium-term variations in $\dot M$,
on scales of hundreds to thousands of years, would manifest as
object-to-object scatter in $T_{\rm eff}$ at a given $P_{\rm orb}$.  In
contrast, the measurements of $T_{\rm eff}$ are remarkably consistent for the
groups.  The novalikes show the best evidence of scatter, and therefore of
$\dot M$ variations on these timescales. They also would have the shortest
adjustment time for their $T_{\rm eff}$ due to the large implied $\timav$ and
luminosities.
Thus novalikes displaying low states might accrete at $\simeq 10^{-8}M_\odot$
yr$^{-1}$ for periods of 100 years or so, and but have $\timav$ of more like a
few $10^{-9}M_\odot$ yr$^{-1}$.  This can be better quantified as more
measurements become available.

The consistency of the $T_{\rm eff}$ values at a given $P_{\rm
orb}$ within each group provides good evidence against large,
medium-term variations in $\dot M$.  For the $\timav$ appropriate for
objects with $T_{\rm eff} \lesssim 20$kK, this statement extends to
even $10^5$ years or more.  That is, the objects for which we have
good measurements of $T_{\rm eff}$ appear to have remarkably similar
$\dot M$ histories over the last $10^3$ to $10^5$ years, though these
objects were likely to have been born at a variety of $P_{\rm orb}$'s.
As noted earlier, a true scatter is expected due to variations in
$M_{\rm acc}$ and $M$ among observed systems. This makes the
consistency among the measurements even more remarkable; enough to
suggest there might be some mechanism, possibly in Classical Nova
outbursts, which regulates $M$ beyond just selection bias.  More
careful treatment of sample bias, in particular with respect to the WD
mass, which actually entails more uniform UV study of known systems,
would be necessary to make conclusions on the presence or absence of
such a mechanism.

The extreme case of uncertainty due to long term $\dot M$ variations
arises from the fact that inactive systems, as e.g. in the hibernation scenario of
\citet{1986ApJ...311..163S} or in irradiation-induced mass transfer
cycles \citep{ritteretal00-1}, may not be
included in the CV census at all, and therefore there is no
opportunity to measure their $T_{\rm eff}$.  This can be roughly
quantified by considering a system with a duty cycle $f$, which is
active for a time period $t_{\rm active}$ and then ceases mass
transfer for a period of $t_{\rm active}(1-f)/f$.  During the active
phase, the $L_{\rm q}$ of such a system is given approximately by
\begin{equation}
L_{\rm q,active} \approx L_{\rm q}(\timav) + \delta L_{\rm q}(\dot M_{\rm active}, t_{\rm
active})
\end{equation}
where the dependence of $\delta L_{\rm q}$ on $t_{\rm active}$ is the same as
the dependence on thermal time shown in Figure \ref{fig:var-time}.  We
introduce a fractional response function $R_L$ such that $\delta L_{\rm
q}(\dot M_{\rm active}, t_{\rm active}) = L_{\rm q}(\dot M_{\rm active})\cdot
R_L(t_{\rm active})$ such that $0\le R_L\le 1$.
Then $R_L(t_{\rm active})$ is the
fractional response to an order unity variation in $\dot M$ on a timescale
$t_{\rm active}$, which increases with $t_{\rm active}$, and is the unitless
quantity actually shown in Figure \ref{fig:var-time}.
Noting that $\dot M_{\rm active}\approx \timav / f$ and
that $L_{\rm q}\propto \dot M$, we have \begin{equation} L_{\rm q,active}
\approx L_{\rm q}(\timav)\cdot(1+R_L(t_{\rm active})/f)\ .  \end{equation} This
demonstrates that for short $t_{\rm active}$, such that $R_L\ll 1$, $L_{\rm
q,active}$ provides a good proxy for $L_{\rm q}(\timav)$, and therefore
$T_{\rm eff}$ is a good indicator of $\timav$ even with a direct conversion.
However, for $R_L(t_{\rm active})\sim 1$ and small $f$, $L_{\rm q}$ during the
active state can be largely unrelated to $\timav$, and instead be set by
$f\approx \timav / \dot M_{\rm active}$ and $t_{\rm active}$.  It should be
noted that a long $t_{\rm active}$ and a low $f$ can imply a very long
recurrence time.

The low duty cycle and long recurrence time scenario just described is very
unlikely to apply to systems with $T_{\rm eff} \lesssim 20$kK.  The proximity
of the $\timav$ implied for these systems to the lower limit set by
gravitational radiation angular momentum losses (dashed lines in figure
\ref{fig:Teff-Porb_wide}) excludes $f\ll 1$.  This means that time series
like those presented in Section \ref{sec:simulations}, where $f=0.5$, are
appropriate for these systems.  Additionally, as discussed above, very long
$t_{\rm active}\gtrsim 10^6$ years would be necessary to ensure consistency
among so many independent objects.

The situation for non-magnetic systems above the period gap is less
constraining.  Assuming that there is some mechanism which can
regulate $\dot M_{\rm active}$ with some precision, the consistency of
several measurements in this region implies that $t_{\rm active}$ must
at least be a few times the time it takes to reach $L_{\rm q}(\dot
M_{\rm active})$.  From the higher $\timav$ curve in Figure
\ref{fig:var-time}, it will take approximately 5,000 years to rise
within 20\% of the $L_{\rm q}(\dot M_{\rm active})$ indicated for
these systems, so that we can estimate $t_{\rm active} \gtrsim 10^4$
years.  In this case we are assuming that $R_L(t_{\rm active})\simeq 1$,
so that in order to overestimate the $\timav$ by a factor of $10$
requires $f=1/10$, and thus a recurrence time of $10^5$ years and a
similar duration of the inactive phase.  Thus long-timescale
hibernation scenarios \citep[e.g.][]{1986ApJ...311..163S}
cannot be excluded by the current $T_{\rm eff}$ data.  There is also
no apparent evidence favoring such scenarios, in the form of downward scatter
of objects still transiting between inactive and active phases.

A sample of $T_{\rm eff}$ measurements for detached systems in the 3-6
hour $P_{\rm orb}$ interval could conclusively rule this out by the
absence of excess WDs with $T_{\rm eff}\simeq 15$~kK compared to longer
periods.  Such samples are
currently being constructed and the results are ambiguous.
Several detached
WD+MS systems that bear the characteristics of hibernating CVs have
been identified, namely BPM\,71214 with $\Porb=290$\,min and
$\Teff=17\,000$\,K, \citep{kawkaetal02-1, kawka+vennes03-1},
EC\,13471--1258 with $\Porb=217$\,min and $\Teff=14\,220\pm350$\,K
\citep{odonoghueetal03-1}, and HS\,2237+8154 with $\Porb=178$\,min and
$\Teff=11\,500$\,K \citep{gaensickeetal04-1}.
The first two show promise, however there is significant selection bias
toward finding hot WDs and there is an expected (contaminant) population of
systems which are just coming into contact and simply have young WD
primaries.

\subsubsection{Inaccessible Quiescent $T_{\rm eff}$ Values}

The second major source of uncertainty in drawing conclusions from the
available set of $T_{\rm eff}$ measurements is due to the set of
circumstances which must come to pass in order to allow direct measurement of
the WD photosphere.  For Dwarf Novae, which accrete in bursts with $\dot M
\gg \timav$, it is also necessary to wait a sufficient period of time after
the outburst in order to get a good idea of the baseline quiescent $T_{\rm
eff}$ that is escaping from the deeper layers of the envelope with longer
thermal times.  As mentioned above, this latter can be achieved largely
empirically by selecting the timing of $T_{\rm eff}$ measurements with
respect to disk outbursts.

Non-magnetic Cataclysmic variables with $P_{\rm orb} < 2$ hours are
predominantly Dwarf Novae \citep{ritter+kolb03-1}. The
selection criteria for such systems are important: the emission of
unknown origin discussed in section \ref{sec:dn} must
be less bright than the WD in the UV, and the absorption of the system
must be low enough that the WD can be measured well, placing a
constraint on the inclination of the system.  Without better
characterization of the unidentified broad-band emission, our only
option is to assume it is a random contaminant which is uncorrelated
with $\timav$.  Again, it is possible to confirm this with better UV
study of known systems.  Since both of these selection criteria are not
expected to correlate with $\timav$, we believe that our sample of
non-magnetic systems in the $P_{\rm orb}< 2$ hour range should be
representative of the typical $\timav$ in these systems.  There might
be a slight bias toward low $\timav$ due to their having more
accessible quiescent intervals, but there are no indications that this
is the case.

For $P_{\rm orb} > 3$~hours, DN are a minority of the population, and thus
there is concern that only particular systems with low $\timav$ have made it
into our sample.  If true, this would imply that the region at higher $T_{\rm
eff}$ than the measured systems in the 3.5-5 hour period range in
Figure~\ref{fig:Teff-Porb_wide} should contain the higher $\timav$ systems
which did not enter our sample.  This would imply a higher $\timav$ than we
can infer directly from the measured systems, and thus impart more favor to
the traditional magnetic braking prescriptions.

\subsection{VY\,Sculptoris stars}
The three VY\,Scl stars with well-determined temperatures stand out in
the 3--4\,h period range containing the hottest CVWDs known, and
consequently have very high mass transfer rates. In effect, the
deduced mass transfer rates exceed those predicted by the standard
evolution theory for the majority of CVs within that period range
\citep{Kolb93, Howeetal01}. $\Teff$ measurements of VY\,Scl require
the fairly prompt observational attention once they enter a low state,
preferably with an UV facility, which explains the small number of
available values. Practically all VY\,Scl stars are located
within the 3--4\,h orbital period range
\citep[e.g.][]{honeycutt+kafka04-1}. Furthermore,
\citet{rodriguez-giletal07-2} have shown that the SW\,Sex stars,
intrinsically bright novalike variables which are intimately related
to the VY\,Scl stars (\citealt{hameury+lasota02-1};
\citealt{hellier00-1}; in fact, the two groups overlap to a large
extent) are the dominant population of CVs in the 3--4\,h orbital
period range. Speculating that high $\Teff$ and $\timav$ are a common
characteristic to all VY\,Scl/SW\,Sex stars suggests that these
systems represent an exceptional phase in CV evolution. One possible
explanation is that these are systems that just evolved into a
semi-detached configuration, as the the mass transfer goes through a
short peak during turn-on \citep[e.g.][]{dantonaetal89-1}, and that
CVs are preferentially born within the 3--4\,h period range, which
would be the case if the initial mass distribution is peaked towards
equal masses in the progenitor main-sequence binaries
\citep{dekool92-1}.

\subsection{Polars and Wind Braking}

In agreement with the interrupted magnetic braking scenario for non-magnetic
CV evolution, these $T_{\rm eff}$ measurements indicate that $\timav$ is
approximately an order of magnitude larger above the period gap than below.
But $T_{\rm eff}$ measurements can do better than this.  Knowing the
relation between $T_{\rm eff}$ and $\timav$ we can say that the
objects below
the gap are roughly consistent with gravitational radiation losses, with
possibly some enhancement of a factor of 2 or 3 for non-magnetic systems.
In contrast $\timav$ above
the period gap is an order of magnitude greater than that predicted by
gravitational radiation alone.  There is, finally, an additional constraint
that is entirely specific to the magnetic braking mechanism: we find a marked
difference between the $\timav$ implied for non-magnetic systems and those of
strongly magnetic systems (Polars) and, as shown in
Figure~\ref{fig:Teff-Porb_wide}, the $\timav$ in Polars is consistent with
that expected from gravitational radiation angular momentum loss alone, while
that in non-magnetic systems is at least an order of magnitude higher.  As
shown in section~\ref{sec:polar} such a decrement in $T_{\rm eff}$, which is
measured away from the poles where the accretion impacts, is too large
to be explained by accretion geometry for the magnetic fields observed.
Additionally, from the discussion in section~\ref{sec:longmdotvary}, such a
large difference would require an extreme assumption about the duty cycle in
non-magnetic systems. Therefore the contrast between magnetic and
non-magnetic systems must arise from a difference in $\timav$.

The lack of an enhanced $\timav$ in magnetic systems arises from changes in
the magnetic field structure near the secondary which hinders the loss of
angular momentum via a wind \citep{LiWuWick94a}.  Our measurements provide
the best direct evidence that this does occur, and additionally that the
resulting $\timav$ is consistent with gravitational radiation.  This provides
very strong support for the basic picture of magnetic braking, though the
precise mechanism by which it ceases is still somewhat mysterious.  We should
highlight that this reduction of magnetic braking in Polars is widely
expected, and was originally proposed to explain the lack of an apparent
period gap in magnetic CVs
\citep{wickramasinghe+wu94-1,lietal94-1,webbink+wickramasinghe02-1}.  The
resulting slower evolution of magnetic systems will also enhance the number
of magnetic CVs relative to nonmagnetic ones with respect to field WDs,
giving roughly the fraction observed \citep{TownBild05}.

\subsection{Below the Period Gap}

\begin{figure}
%\plotone{Teff-Porb.eps}
%\includegraphics[width=\columnwidth]{Teff-Porb.eps}
\includegraphics[width=\columnwidth]{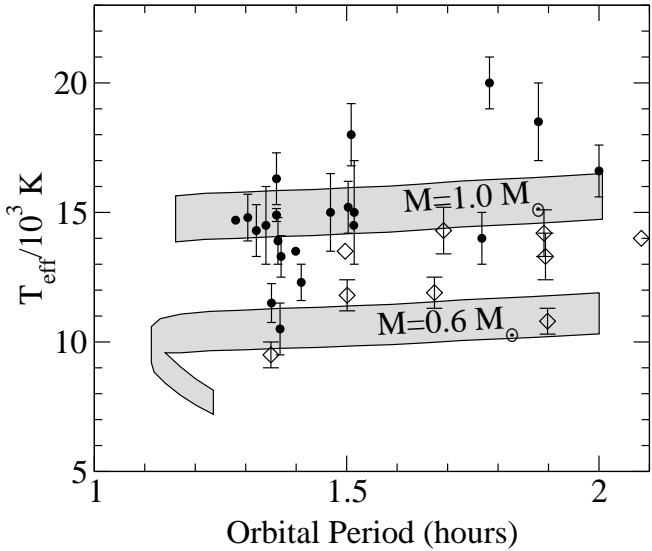}
\caption{
\label{fig:Teff-Porb}
Closeup of $P_{\rm orb} < 2$ hours, filled circles indicate non-magnetic
systems and open diamonds indicate magnetic systems (see Figure
\ref{fig:Teff-Porb_wide}).  Overplotted are the range of $T_{\rm
eff}$ expected for accretion driven by gravitational radiation
\citep{kolb+baraffe99-1} for two WD masses.  Note that the theoretical
minimum period is notably shorter than that inferred from the data.
}
\end{figure}

Due to the need for low $\timav$, many of of the high-quality $T_{\rm
eff}$ measurements are for systems with $P_{\rm orb} < 2$~hours.  An
expanded plot of this region is shown in Figure~\ref{fig:Teff-Porb}.  For
$M=0.6M_\odot$ we use the $\timav$ from \citet{kolb+baraffe99-1} directly
and for $1.0M_\odot$ we use their $M$-$R$ relation for the secondary
and gravitational radiation angular momentum losses. We display the
predicted region for $0.05M_{\rm ign}<M_{\rm acc} < 0.95 M_{\rm ign}$
which object will traverse during buildup toward a classical nova when
$M_{\rm acc}=M_{\rm ign}$.  Many of the measurements for non-magnetic
objects are clustered about $T_{\rm eff}= 15$~kK, providing evidence,
as discussed above, against long-term $\dot M$ variations.  The
exception to this is that there is significant downward scatter at the
shortest orbital periods 1.3-1.4 hours. The presence of this scatter
at these $P_{\rm orb}$ and not at slightly longer ones strongly
suggests that the $\timav$ in these objects has begun to decline as
expected when they pass beyond period minimum \citep{Howeetal01}.
This feature highlights the disagreement of the observed period
minimum and that predicted by theory \citep{kolb+baraffe99-1},
which is apparent in Figure~\ref{fig:Teff-Porb} because the minimum is
not very sensitive to $M$.  As discussed in \citet{TownBild03} we find
that either short-period CVs have $M\simeq 0.9$-$1.0M_\odot$
(such as e.g. found in SDSS\,1035+0555,
\citet{littlefairetal06-2}) or mass transfer is enhanced by a factor
of 2-3 over that predicted by only gravitational radiation losses.
There is also some evidence from the few measurements near $P_{\rm
orb}=1.8$ hours that the dependence of $\timav$ on $P_{\rm orb}$ is
not as flat as is predicted by gravitational radiation.  More
measurements will be necessary before conclusions can be drawn.

Even here below the period gap there is a smaller but significant
difference between strongly magnetic and non-magnetic systems.  The
energy released by compression deeper than the point at which the
material is able to spread over the star amounts to better than 80\%
of the $L_{\rm q}$ expected in the non-magnetic case.  However such a
decrease in $L_{\rm q}$ only reduces $T_{\rm eff}$ from 15 kK to 14
kK, not enough to account for the difference in $T_{\rm eff}$ between
magnetic and non-magnetic systems below the gap.  For high-field
cases, $B\simeq 10^8$ G, the reduction can reach 50\%, bringing
$T_{\rm eff}$ down to 13 kK.  However, none of the systems used
in this paper (Table\,\ref{t-teff}) has such a high field (see
Sect.\,\ref{sec:am}).  It still appears that either $\timav$ is
slightly above that due to gravitational radiation or these
magnetic objects do have typical $M\simeq 0.7M_\odot$.  A
typical mass above $0.6M_\odot$ is expected both from simple selection
bias because the luminosity increases with $M$ and since
field magnetic WDs tend to be higher mass than non-magnetics.

\acknowledgements

The authors would like to thank Lars Bildsten and Christian Knigge for
stimulating discussions.  We also thank the anonymous referee for
insightful comments.  This work was supported by the National
Science Foundation under grants PHY99-07949, and AST02-05956, and by NASA
through grant AR-09517.01-A from STScI, which is operated by AURA, Inc.,
under NASA contract NAS5-26555.  D.M.T. is supported by the NSF Physics
Frontier Centers' Joint Institute for Nuclear Astrophysics under grant PHY
02-16783 and DOE under grant DE-FG 02-91ER 40606.

\appendix

\section{Simulation of hydrostatic envelope in plane parallel}

In this appendix we detail our treatment via simulation of the envelope of
the accreting WD.  We work in the plane parallel, hydrostatic approximation
so that $P = gy$ at every point.  Discretizing the temperature in $y$ and
time, the heat equation, (\ref{eq:heat}), is integrated forward in time using
a Crank-Nicholson like integration rule
\begin{equation}
\label{eq:crank}
\frac{T^{n+1}_k-T^{n}_k}{t_{n+1}-t_n} = \frac12\left[\left.\frac{\partial
T}{\partial t}\right|_{t_n,y_k}
+\left.\pder{T}{t}\right|_{t_{n+1},y_k} \right]
\end{equation}
where the derivatives of $T$ in $y$ on the grid are evaluated via centered differences
\begin{equation}
\left.\pder{T}{y}\right|_{y_k} = \frac{T_{k+1}-T_{k-1}}{y_{k+1}-y_{k-1}},
\quad\quad
\left.\pder{^2T}{y^2}\right|_{y_k}
= \frac{T_{k+1}-T_{k}}{y_{k+1}-y_{k}}
-\frac{T_{k}-T_{k-1}}{y_{k}-y_{k-1}}.
\end{equation}
Initial conditions are taken from the static models used by TB04.  Note that
eq. (\ref{eq:crank}) is an implicit integration rule, representing $N$
equations in $N$ unknowns, which is solved for $T_k^{n+1}$ using a
Newton-Raphson iteration.  Timesteps are chosen such that the maximum $\Delta
T/T$ at any point on the grid is less that $10^{-3}$.  The steady state
solution produced by the time-dependent code compares very well with the static
solutions produced by direct integration of the structure equations in TB04.

The flux at the outer boundary is found by integrating eq. (\ref{eq:heat}) in
from the photosphere using a 4th order Runge-Kutta integrator.  A shooting
method (root-find) is used to vary $F_{\rm photosphere}$ to match $T(y_{\rm
outer})$ at the outer edge of the simulation grid, $y_{\rm outer}$, and the
resulting $F(y_{\rm outer})$ is used for the flux at the edge of the grid.
While there are no convection zones in the simulation domain, there is a
convection zone in the part of the envelope which forms the boundary
condition.  Convection here is treated with the ML2 formalism \citep{ML2}.
Use of this boundary condition is equivalent to the approximation that the
thermal time of the layer of depth $y_{\rm outer}$ is negligible.
The inner boundary condition is constant temperature, fixed to the
equilibrium temperature at the $\timav$ under consideration (TB04).

Two different grids are used in this work.  For the long-timescale variation
studies of section \ref{sec:simulations}, we use $N=120$ points evenly spaced in
$\log y$ extending from from $y=6\times 10^5$ g cm$^{-2}$ to $3.9\times
10^{12}$ g cm$^{-2}$, corresponding to mass coordinates of $1.6\times
10^{-9}M_\odot$ and $0.01M_\odot$, respectively on a $0.9M_\odot$ WD,
$R=6.45\times 10^8$ cm.  For $\timav = 5\times 10^{-11}M_\odot$ yr$^{-1}$ at
this mass, the thermal timescale of the outermost point $t_{\rm th}(y_{\rm
outer})\simeq10$ days, much shorter than any variability being studied.  For
the study of TT Ari in section \ref{sec:system}, the upper edge of the grid
was extended to shorter thermal timescales.  In this case we used 100 points
between $y=2.75\times10^4$ and $3.64\times 10^8$~g~cm$^{-2}$ and 50 points
extending down to $3.9\times 10^{12}$~g~cm$^{-2}$, for a total 150. This
outer point corresponds to $t_{\rm th}(y_{\rm outer})\simeq 15$ hours for the
parameters used to reproduce TT Ari.

The depth of the boundary between the solar abundance accreted material and
the underlying 50/50 carbon and oxygen material, $y_{\rm HHe}=M_{\rm
acc}/(4\pi R^2)$ is tracked with a separate variable whose evolution is
directly specified from $\dot M(t)$.  Although selecting between two
compositions depending upon whether $y_k<y_{\rm HHe}$ or $y_{\rm HHe}<y_k$
gives suitably accurate results, a smooth progression is significantly more
easily integrated numerically.  This allows a smooth change in abundance at a
given point while still allowing a sharp interface to be represented on a
modest resolution grid.  We now proceed to describe our interface treatment
in detail.

Consider the interface as being at column depth $y_I=y_{\rm HHe}$ lying
between two gridpoints at depths $y_{k}$ and $y_{k+1}$.  The temperature at
each of these points is $T_{k}$ and $T_{k+1}$ respectively, and that at the
interface is $T_I$.  Both the temperature and flux must be continuous at the
interface, so letting $F_H$ and $F_C$ indicate the flux evaluated at $y_I$ on
the H/He side and the C/O side respectively, we have
\begin{equation}
\label{eq:fluxrelation}
F_I = \frac{ 4acT^3}{3\kappa}\pder{T}{y} = F_H = F_C \quad{\rm or}\quad
\frac{1}{\kappa_H(y_I,T_I)}\left(\pder{T}{y}\right)_{H}
=\frac{1}{\kappa_C(y_I,T_I)}\left(\pder{T}{y}\right)_{C}
\end{equation}
where the subscripts $H$ and $C$ on the derivatives indicate evaluation on
respective sides of the interface and on $\kappa$ indicate evaluation at
$P_I$ and $T_I$ but with H/He and C/O composition respectively.
It is inadvisable to attempt to solve for $T_I$ directly because $y_I-y_k$
can become arbitrarily small and lead to singularities.  Instead we will
solve for the derivatives.  The interface temperature can be written by
expanding from both directions
\begin{equation}
\label{eq:texpress}
T_I = T_k + (y_I-y_k)\left(\pder{T}{y}\right)_H
= T_{k+1} + (y_I-y_{k+1})\left(\pder{T}{y}\right)_C
\end{equation}
which upon combination with eq. (\ref{eq:fluxrelation}) gives
\begin{equation}
0=
T_{k+1} -T_k+ 
\left[(y_I-y_{k+1})\frac{\kappa_C(T_I)}{\kappa_H(T_I)}
 - (y_I-y_k)\right]\left(\pder{T}{y}\right)_H\ .
\end{equation}
By using the first expression for $T_{I}$ from (\ref{eq:texpress}), this
can be solved for $(\partial T/\partial y)_{H}$.

Finally, all the derivatives near the interface can then be constructed from
the grid quantities and $(\partial T/\partial y)_H$ in a way which mimics
centered differencing.  We let $(\pder{T}{y})_H$ stand in for the first order
difference at the midpoint between $k$ and $k+1$, since it should be
approximately what that first order difference would have been if there
were no change in composition. This gives
\begin{eqnarray}
\left(\der{T}{y}\right)_k &=&
\frac12\left[
\left(\pder{T}{y}\right)_H
+\frac{T_k-T_{k-1}}{y_k-y_{k-1}}
\right]\ ,\\
\left(\der{^2T}{y^2}\right)_k &=&
\frac{2}{y_{k+1}-y_{k-1}}\left[
\left(\pder{T}{y}\right)_H
-\frac{T_k-T_{k-1}}{y_k-y_{k-1}}\right]\ ,\\
\left(\der{T}{y}\right)_{k+1} &=&
\frac12\left[
\frac{T_{k+2}-T_{k+1}}{y_{k+2}-y_{k+1}}
+ \left(\pder{T}{y}\right)_C
\right]\ , \\
\left(\der{^2T}{y^2}\right)_{k+1} &=&
\frac{2}{y_{k+2}-y_{k}}\left[
\frac{T_{k+2}-T_{k+1}}{y_{k+2}-y_{k+1}}
- \left(\pder{T}{y}\right)_C
\right]\ .
\end{eqnarray}

For the outer portions of the envelope we use the 2002 update to the OPAL
equation of state tables \citep{RogeSwenIgle96}, and for higher densities we
use the analytical approximations for a fully ionized plasma from
\citet{Pacz83} and Coulomb correction from \citet{ChabPote98}.  While these
two EOS methods are very consistent at the table edge, a linear average in a
crossover region of a factor of 5 in density is used to smooth the boundary.
OPAL radiative opacities \citep{IgleRoge96} are also used along with
conductivities from \citet{Itohetal83}.

\bibliography{apj-jour,T2,aabib,proceedings,submitted}

\begin{thebibliography}{145}
\expandafter\ifx\csname natexlab\endcsname\relax\def\natexlab#1{#1}\fi

\bibitem[{{Andronov} {et~al.}(2003){Andronov}, {Pinsonneault}, \&
  {Sills}}]{Andretal03}
{Andronov}, N., {Pinsonneault}, M., \& {Sills}, A. 2003, \apj, 582, 358

\bibitem[{{Araujo-Betancor} {et~al.}(2005{\natexlab{a}}){Araujo-Betancor},
  {G{\" a}nsicke}, {Hagen}, {Marsh}, {Harlaftis}, {Thorstensen}, {Fried},
  {Schmeer}, \& {Engels}}]{araujo-betancoretal05-1}
{Araujo-Betancor}, S., {G{\" a}nsicke}, B.~T., {Hagen}, H.-J., {Marsh}, T.~R.,
  {Harlaftis}, E.~T., {Thorstensen}, J., {Fried}, R.~E., {Schmeer}, P., \&
  {Engels}, D. 2005{\natexlab{a}}, \aap, 430, 629

\bibitem[{{Araujo-Betancor} {et~al.}(2005{\natexlab{b}}){Araujo-Betancor},
  {G{\" a}nsicke}, {Long}, {Beuermann}, {de Martino}, {Sion}, \&
  {Szkody}}]{araujo-betancoretal05-2}
{Araujo-Betancor}, S., {G{\" a}nsicke}, B.~T., {Long}, K.~S., {Beuermann}, K.,
  {de Martino}, D., {Sion}, E.~M., \& {Szkody}, P. 2005{\natexlab{b}}, \apj,
  622, 589

\bibitem[{{Araujo-Betancor} {et~al.}(2003){Araujo-Betancor}, {Knigge}, {Long},
  {Hoard}, {Szkody}, {Rodgers}, {Krisciunas}, {Dhillon}, {Hynes}, {Patterson},
  \& {Kemp}}]{araujo-betancoretal03-1}
{Araujo-Betancor}, S., {Knigge}, C., {Long}, K.~S., {Hoard}, D.~W., {Szkody},
  P., {Rodgers}, B., {Krisciunas}, K., {Dhillon}, V.~S., {Hynes}, R.~I.,
  {Patterson}, J., \& {Kemp}, J. 2003, \apj, 583, 437

\bibitem[{{Belle} {et~al.}(2003){Belle}, {Howell}, {Sion}, {Long}, \&
  {Szkody}}]{belleetal03-1}
{Belle}, K.~E., {Howell}, S.~B., {Sion}, E.~M., {Long}, K.~S., \& {Szkody}, P.
  2003, \apj, 587, 373

\bibitem[{{Bergeron} {et~al.}(1992){Bergeron}, {Wesemael}, \& {Fontaine}}]{ML2}
{Bergeron}, P., {Wesemael}, F., \& {Fontaine}, G. 1992, \apj, 387, 288

\bibitem[{{Beuermann}(2006)}]{Beue06}
{Beuermann}, K. 2006, \aap, 460, 783

\bibitem[{{Beuermann} {et~al.}(1998){Beuermann}, {Baraffe}, {Kolb}, \&
  {Weichhold}}]{beuermannetal98-1}
{Beuermann}, K., {Baraffe}, I., {Kolb}, U., \& {Weichhold}, M. 1998, \aap, 339,
  518

\bibitem[{{Beuermann} {et~al.}(2004){Beuermann}, {Harrison}, {McArthur},
  {Benedict}, \& {G{\" a}nsicke}}]{beuermannetal04-1}
{Beuermann}, K., {Harrison}, T.~E., {McArthur}, B.~E., {Benedict}, G.~F., \&
  {G{\" a}nsicke}, B.~T. 2004, \aap, 419, 291

\bibitem[{{Beuermann} {et~al.}(2000){Beuermann}, {Wheatley}, {Ramsay},
  {Euchner}, \& {G\"ansicke}}]{beuermannetal00-1}
{Beuermann}, K., {Wheatley}, P., {Ramsay}, G., {Euchner}, F., \& {G\"ansicke},
  B.~T. 2000, \aap, 354, L49

\bibitem[{{Chabrier} \& {Potekhin}(1998)}]{ChabPote98}
{Chabrier}, G. \& {Potekhin}, A.~Y. 1998, \pre, 58, 4941

\bibitem[{{Cheng} {et~al.}(2000){Cheng}, {Horne}, {Marsh}, {Hubeny}, \&
  {Sion}}]{chengetal00-1}
{Cheng}, F.~H., {Horne}, K., {Marsh}, T.~R., {Hubeny}, I., \& {Sion}, E.~M.
  2000, \apj, 542, 1064

\bibitem[{{D'Antona} {et~al.}(1989){D'Antona}, {Mazzitelli}, \&
  {Ritter}}]{dantonaetal89-1}
{D'Antona}, F., {Mazzitelli}, I., \& {Ritter}, H. 1989, \aap, 225, 391

\bibitem[{{de Kool}(1992)}]{dekool92-1}
{de Kool}, M. 1992, \aap, 261, 188

\bibitem[{{de Martino} {et~al.}(2006){de Martino}, {Bonnet-Bidaud}, {Mouchet},
  {G{\"a}nsicke}, {Haberl}, \& {Motch}}]{demartinoetal06-1}
{de Martino}, D., {Bonnet-Bidaud}, J.-M., {Mouchet}, M., {G{\"a}nsicke}, B.~T.,
  {Haberl}, F., \& {Motch}, C. 2006, \aap, 449, 1151

\bibitem[{{Dhillon} {et~al.}(2007){Dhillon}, {Marsh}, {Stevenson}, {Atkinson},
  {Kerry}, {Peacocke}, {Vick}, {Beard}, {Ives}, {Lunney}, {McLay}, {Tierney},
  {Kelly}, {Littlefair}, {Nicholson}, {Pashley}, {Harlaftis}, \&
  {O'Brien}}]{dhillonetal07-1}
{Dhillon}, V.~S., {Marsh}, T.~R., {Stevenson}, M.~J., {Atkinson}, D.~C.,
  {Kerry}, P., {Peacocke}, P.~T., {Vick}, A.~J.~A., {Beard}, S.~M., {Ives},
  D.~J., {Lunney}, D.~W., {McLay}, S.~A., {Tierney}, C.~J., {Kelly}, J.,
  {Littlefair}, S.~P., {Nicholson}, R., {Pashley}, R., {Harlaftis}, E.~T., \&
  {O'Brien}, K. 2007, \mnras, 378, 825

\bibitem[{{Eisenbart} {et~al.}(2002){Eisenbart}, {Beuermann}, {Reinsch}, \&
  {G{\" a}nsicke}}]{eisenbartetal02-1}
{Eisenbart}, S., {Beuermann}, K., {Reinsch}, K., \& {G{\" a}nsicke}, B.~T.
  2002, \aap, 382, 984

\bibitem[{{Epelstain} {et~al.}(2007){Epelstain}, {Yaron}, {Kovetz}, \&
  {Prialnik}}]{Epeletal07}
{Epelstain}, N., {Yaron}, O., {Kovetz}, A., \& {Prialnik}, D. 2007, \mnras,
  374, 1449

\bibitem[{{Eracleous} {et~al.}(1994){Eracleous}, {Horne}, {Robinson}, {Zhang},
  {Marsh}, \& {Wood}}]{eracleousetal94-1}
{Eracleous}, M., {Horne}, K., {Robinson}, E.~L., {Zhang}, E.-H., {Marsh},
  T.~R., \& {Wood}, J.~H. 1994, \apj, 433, 313

\bibitem[{{Faulkner}(1971)}]{Faul71}
{Faulkner}, J. 1971, \apjl, 170, L99+

\bibitem[{{Feline} {et~al.}(2004){Feline}, {Dhillon}, {Marsh}, {Stevenson},
  {Watson}, \& {Brinkworth}}]{felineetal04-1}
{Feline}, W.~J., {Dhillon}, V.~S., {Marsh}, T.~R., {Stevenson}, M.~J.,
  {Watson}, C.~A., \& {Brinkworth}, C.~S. 2004, \mnras, 347, 1173

\bibitem[{{Feline} {et~al.}(2005){Feline}, {Dhillon}, {Marsh}, {Watson}, \&
  {Littlefair}}]{felineetal05-1}
{Feline}, W.~J., {Dhillon}, V.~S., {Marsh}, T.~R., {Watson}, C.~A., \&
  {Littlefair}, S.~P. 2005, \mnras, 364, 1158

\bibitem[{{G{\" a}nsicke} {et~al.}(2001){G{\" a}nsicke}, {Schmidt}, {Jordan},
  \& {Szkody}}]{GansSchm01}
{G{\" a}nsicke}, B.~T., {Schmidt}, G.~D., {Jordan}, S., \& {Szkody}, P. 2001,
  \apj, 555, 380

\bibitem[{{G{\" a}nsicke} {et~al.}(2005){G{\" a}nsicke}, {Szkody}, {Howell}, \&
  {Sion}}]{gaensickeetal05-2}
{G{\" a}nsicke}, B.~T., {Szkody}, P., {Howell}, S.~B., \& {Sion}, E.~M. 2005,
  \apj, 629, 451

\bibitem[{{G\"ansicke}(1999)}]{gaensicke99-1}
{G\"ansicke}, B.~T. 1999, in Annapolis Workshop on Magnetic Cataclysmic
  Variables, ed. C.~{Hellier} \& K.~{Mukai} (ASP Conf. Ser. 157), 261--272

\bibitem[{{G\"ansicke} {et~al.}(2004{\natexlab{a}}){G\"ansicke},
  {Araujo-Betancor}, {Hagen}, {Harlaftis}, {Kitsionas}, {Dreizler}, \&
  {Engels}}]{gaensickeetal04-1}
{G\"ansicke}, B.~T., {Araujo-Betancor}, S., {Hagen}, H.-J., {Harlaftis}, E.~T.,
  {Kitsionas}, S., {Dreizler}, S., \& {Engels}, D. 2004{\natexlab{a}}, \aap,
  418, 265

\bibitem[{{G\"ansicke} \&
  {Beuermann}(1996{\natexlab{a}})}]{gaensicke+beuermann96-3}
{G\"ansicke}, B.~T. \& {Beuermann}, K. 1996{\natexlab{a}}, in
  R\"ontgenstrahlung from the Universe, ed. H.~U. {Zimmermann}, J.~{Tr\"umper},
  \& H.~{Yorke}, MPE Report No. 263 (Garching: MPE), 137--138

\bibitem[{{G\"ansicke} \&
  {Beuermann}(1996{\natexlab{b}})}]{gaensicke+beuermann96-1}
{G\"ansicke}, B.~T. \& {Beuermann}, K. 1996{\natexlab{b}}, \aap, 309, L47

\bibitem[{{G\"ansicke} {et~al.}(1995){G\"ansicke}, {Beuermann}, \& {de
  Martino}}]{gaensickeetal95-1}
{G\"ansicke}, B.~T., {Beuermann}, K., \& {de Martino}, D. 1995, \aap, 303, 127

\bibitem[{{G\"ansicke} {et~al.}(2000){G\"ansicke}, {Beuermann}, {de Martino},
  \& {Thomas}}]{gaensickeetal00-1}
{G\"ansicke}, B.~T., {Beuermann}, K., {de Martino}, D., \& {Thomas}, H.-C.
  2000, \aap, 354, 605

\bibitem[{{G\"ansicke} {et~al.}(2004{\natexlab{b}}){G\"ansicke}, {Jordan},
  {Beuermann}, {de Martino}, {Szkody}, {Marsh}, \&
  {Thorstensen}}]{gaensickeetal04-3}
{G\"ansicke}, B.~T., {Jordan}, S., {Beuermann}, K., {de Martino}, D., {Szkody},
  P., {Marsh}, T.~R., \& {Thorstensen}, J. 2004{\natexlab{b}}, \apjl, 613, L141

\bibitem[{{G\"ansicke} \& {Koester}(1999)}]{gaensicke+koester99-1}
{G\"ansicke}, B.~T. \& {Koester}, D. 1999, \aap, 346, 151

\bibitem[{{G{\"a}nsicke} {et~al.}(2006){G{\"a}nsicke}, {Long}, {Barstow}, \&
  {Hubeny}}]{gaensickeetal06-2}
{G{\"a}nsicke}, B.~T., {Long}, K.~S., {Barstow}, M.~A., \& {Hubeny}, I. 2006,
  \apj, 639, 1039

\bibitem[{{G\"ansicke} {et~al.}(1999){G\"ansicke}, {Sion}, {Beuermann},
  {Fabian}, {Cheng}, \& {Krautter}}]{gaensickeetal99-1}
{G\"ansicke}, B.~T., {Sion}, E.~M., {Beuermann}, K., {Fabian}, D., {Cheng},
  F.~H., \& {Krautter}, J. 1999, \aap, 347, 178

\bibitem[{{G\"ansicke} {et~al.}(2001){G\"ansicke}, {Szkody}, {Sion}, {Hoard},
  {Howell}, {Cheng}, \& {Hubeny}}]{gaensickeetal01-3}
{G\"ansicke}, B.~T., {Szkody}, P., {Sion}, E.~M., {Hoard}, D.~W., {Howell}, S.,
  {Cheng}, F.~H., \& {Hubeny}, I. 2001, \aap, 374, 656

\bibitem[{{Godon} {et~al.}(2006){Godon}, {Seward}, {Sion}, \&
  {Szkody}}]{godonetal06-1}
{Godon}, P., {Seward}, L., {Sion}, E.~M., \& {Szkody}, P. 2006, \aj, 131, 2634

\bibitem[{{Godon} {et~al.}(2004){Godon}, {Sion}, {Cheng}, {G{\" a}nsicke},
  {Howell}, {Knigge}, {Sparks}, \& {Starrfield}}]{godonetal04-1}
{Godon}, P., {Sion}, E.~M., {Cheng}, F., {G{\" a}nsicke}, B.~T., {Howell}, S.,
  {Knigge}, C., {Sparks}, W.~M., \& {Starrfield}, S. 2004, \apj, 602, 336

\bibitem[{{Greenstein}(1957)}]{greenstein57-1}
{Greenstein}, J.~L. 1957, \apj, 126, 23

\bibitem[{{Haberl} {et~al.}(2002){Haberl}, {Motch}, \&
  {Zickgraf}}]{haberletal02-1}
{Haberl}, F., {Motch}, C., \& {Zickgraf}, F.-J. 2002, \aap, 387, 201

\bibitem[{{Hameury} {et~al.}(1983){Hameury}, {Bonazzola}, {Heyvaerts}, \&
  {Lasota}}]{Hameetal83}
{Hameury}, J.~M., {Bonazzola}, S., {Heyvaerts}, J., \& {Lasota}, J.~P. 1983,
  \aap, 128, 369

\bibitem[{{Hameury} {et~al.}(1988){Hameury}, {King}, {Lasota}, \&
  {Ritter}}]{Hameetal88}
{Hameury}, J.~M., {King}, A.~R., {Lasota}, J.~P., \& {Ritter}, H. 1988, \mnras,
  231, 535

\bibitem[{{Hameury} \& {Lasota}(2002)}]{hameury+lasota02-1}
{Hameury}, J.~M. \& {Lasota}, J.~P. 2002, \aap, 394, 231

\bibitem[{{Harrison} {et~al.}(2004){Harrison}, {Johnson}, {McArthur},
  {Benedict}, {Szkody}, {Howell}, \& {Gelino}}]{harrisonetal04-3}
{Harrison}, T.~E., {Johnson}, J.~J., {McArthur}, B.~E., {Benedict}, G.~F.,
  {Szkody}, P., {Howell}, S.~B., \& {Gelino}, D.~M. 2004, \aj, 127, 460

\bibitem[{{Hartley} {et~al.}(2005){Hartley}, {Long}, {Froning}, \&
  {Drew}}]{hartleyetal05-1}
{Hartley}, L.~E., {Long}, K.~S., {Froning}, C.~S., \& {Drew}, J.~E. 2005, \apj,
  623, 425

\bibitem[{{Heise} \& {Verbunt}(1988)}]{heise+verbunt88-1}
{Heise}, J. \& {Verbunt}, F. 1988, \aap, 189, 112

\bibitem[{{Hellier}(2000)}]{hellier00-1}
{Hellier}, C. 2000, New Astronomy Review, 44, 131

\bibitem[{{Hessman} {et~al.}(2000){Hessman}, {G\"ansicke}, \&
  {Mattei}}]{hessmanetal00-1}
{Hessman}, F.~V., {G\"ansicke}, B.~T., \& {Mattei}, J.~A. 2000, \aap, 361, 952

\bibitem[{{Hessman} {et~al.}(1989){Hessman}, {Koester}, {Schoembs}, \&
  {Barwig}}]{hessmanetal89-1}
{Hessman}, F.~V., {Koester}, D., {Schoembs}, R., \& {Barwig}, H. 1989, \aap,
  213, 167

\bibitem[{{Hoard} {et~al.}(2004){Hoard}, {Linnell}, {Szkody}, {Fried}, {Sion},
  {Hubeny}, \& {Wolfe}}]{hoardetal04-1}
{Hoard}, D.~W., {Linnell}, A.~P., {Szkody}, P., {Fried}, R.~E., {Sion}, E.~M.,
  {Hubeny}, I., \& {Wolfe}, M.~A. 2004, \apj, 604, 346

\bibitem[{{Honeycutt} \& {Kafka}(2004)}]{honeycutt+kafka04-1}
{Honeycutt}, R.~K. \& {Kafka}, S. 2004, \aj, 128, 1279

\bibitem[{{Horne} {et~al.}(1994){Horne}, {Marsh}, {Cheng}, {Hubeny}, \&
  {Lanz}}]{horneetal94-1}
{Horne}, K., {Marsh}, T.~R., {Cheng}, F.~H., {Hubeny}, I., \& {Lanz}, T. 1994,
  \apj, 426, 294

\bibitem[{{Howell} {et~al.}(2002){Howell}, {G{\" a}nsicke}, {Szkody}, \&
  {Sion}}]{howelletal02-1}
{Howell}, S.~B., {G{\" a}nsicke}, B.~T., {Szkody}, P., \& {Sion}, E.~M. 2002,
  \apj, 575, 419

\bibitem[{{Howell} {et~al.}(2001){Howell}, {Nelson}, \&
  {Rappaport}}]{Howeetal01}
{Howell}, S.~B., {Nelson}, L.~A., \& {Rappaport}, S. 2001, \apj, 550, 897

\bibitem[{{Huang} {et~al.}(1996){Huang}, {Sion}, {Hubeny}, {Cheng}, \&
  {Szkody}}]{huangetal96-1}
{Huang}, M., {Sion}, E.~M., {Hubeny}, I., {Cheng}, F.~H., \& {Szkody}, P. 1996,
  \aj, 111, 2386

\bibitem[{{Iglesias} \& {Rogers}(1996)}]{IgleRoge96}
{Iglesias}, C.~A. \& {Rogers}, F.~J. 1996, \apj, 464, 943

\bibitem[{{Itoh} {et~al.}(1983){Itoh}, {Mitake}, {Iyetomi}, \&
  {Ichimaru}}]{Itohetal83}
{Itoh}, N., {Mitake}, S., {Iyetomi}, H., \& {Ichimaru}, S. 1983, \apj, 273, 774

\bibitem[{{Ivanova} \& {Taam}(2004)}]{IvanTaam04}
{Ivanova}, N. \& {Taam}, R.~E. 2004, \apj, 601, 1058

\bibitem[{{Jordan}(1992)}]{jordan92-1}
{Jordan}, S. 1992, \aap, 265, 570

\bibitem[{{Kawka} \& {Vennes}(2003)}]{kawka+vennes03-1}
{Kawka}, A. \& {Vennes}, S. 2003, \aj, 125, 1444

\bibitem[{{Kawka} {et~al.}(2002){Kawka}, {Vennes}, {Koch}, \&
  {Williams}}]{kawkaetal02-1}
{Kawka}, A., {Vennes}, S., {Koch}, R., \& {Williams}, A. 2002, \aj, 124, 2853

\bibitem[{{Knigge}(2006)}]{knigge06-1}
{Knigge}, C. 2006, \mnras, 373, 484

\bibitem[{{Knigge} {et~al.}(2000){Knigge}, {Long}, {Hoard}, {Szkody}, \&
  {Dhillon}}]{kniggeetal00-1}
{Knigge}, C., {Long}, K.~S., {Hoard}, D.~W., {Szkody}, P., \& {Dhillon}, V.~S.
  2000, \apjl, 539, L49

\bibitem[{{Koester} {et~al.}(1985){Koester}, {Weidemann}, {Zeidler-K.T.}, \&
  {Vauclair}}]{koesteretal85-1}
{Koester}, D., {Weidemann}, V., {Zeidler-K.T.}, E.~M., \& {Vauclair}, G. 1985,
  \aap, 142, L5

\bibitem[{{Kolb}(1993)}]{Kolb93}
{Kolb}, U. 1993, \aap, 271, 149

\bibitem[{{Kolb} \& {Baraffe}(1999)}]{kolb+baraffe99-1}
{Kolb}, U. \& {Baraffe}, I. 1999, \mnras, 309, 1034

\bibitem[{{Li} {et~al.}(1994{\natexlab{a}}){Li}, {Wu}, \&
  {Wickramasinghe}}]{lietal94-1}
{Li}, J.~K., {Wu}, K.~W., \& {Wickramasinghe}, D.~T. 1994{\natexlab{a}},
  \mnras, 270, 769

\bibitem[{{Li} {et~al.}(1994{\natexlab{b}}){Li}, {Wu}, \&
  {Wickramasinghe}}]{LiWuWick94a}
---. 1994{\natexlab{b}}, \mnras, 268, 61

\bibitem[{{Littlefair} {et~al.}(2006{\natexlab{a}}){Littlefair}, {Dhillon},
  {Marsh}, \& {G{\"a}nsicke}}]{littlefairetal06-1}
{Littlefair}, S.~P., {Dhillon}, V.~S., {Marsh}, T.~R., \& {G{\"a}nsicke}, B.~T.
  2006{\natexlab{a}}, \mnras, 371, 1435

\bibitem[{{Littlefair} {et~al.}(2006{\natexlab{b}}){Littlefair}, {Dhillon},
  {Marsh}, {G{\"a}nsicke}, {Southworth}, \& {Watson}}]{littlefairetal06-2}
{Littlefair}, S.~P., {Dhillon}, V.~S., {Marsh}, T.~R., {G{\"a}nsicke}, B.~T.,
  {Southworth}, J., \& {Watson}, C.~A. 2006{\natexlab{b}}, Science, 314, 1578

\bibitem[{{Livio} \& {Pringle}(1994)}]{livio+pringle94-1}
{Livio}, M. \& {Pringle}, J.~E. 1994, \apj, 427, 956

\bibitem[{{Long} {et~al.}(1993){Long}, {Blair}, {Bowers}, {Davidsen}, {Kriss},
  {Sion}, \& {Hubeny}}]{longetal93-1}
{Long}, K.~S., {Blair}, W.~P., {Bowers}, C.~W., {Davidsen}, A.~F., {Kriss},
  G.~A., {Sion}, E.~M., \& {Hubeny}, I. 1993, \apj, 405, 327

\bibitem[{{Long} {et~al.}(1996){Long}, {Blair}, {Hubeny}, \&
  {Raymond}}]{longetal96-2}
{Long}, K.~S., {Blair}, W.~P., {Hubeny}, I., \& {Raymond}, J.~C. 1996, \apj,
  466, 964

\bibitem[{{Long} {et~al.}(2006){Long}, {Brammer}, \& {Froning}}]{longetal06-1}
{Long}, K.~S., {Brammer}, G., \& {Froning}, C.~S. 2006, \apj, 648, 541

\bibitem[{{Long} {et~al.}(2005){Long}, {Froning}, {Knigge}, {Blair}, {Kallman},
  \& {Ko}}]{longetal05-1}
{Long}, K.~S., {Froning}, C.~S., {Knigge}, C., {Blair}, W.~P., {Kallman},
  T.~R., \& {Ko}, Y.-K. 2005, \apj, 630, 511

\bibitem[{{Long} \& {Gilliland}(1999)}]{long+gilliland99-1}
{Long}, K.~S. \& {Gilliland}, R.~L. 1999, \apj, 511, 916

\bibitem[{{Long} {et~al.}(2004){Long}, {Sion}, {G{\" a}nsicke}, \&
  {Szkody}}]{longetal04-1}
{Long}, K.~S., {Sion}, E.~M., {G{\" a}nsicke}, B.~T., \& {Szkody}, P. 2004,
  \apj, 602, 948

\bibitem[{{Mateo} \& {Szkody}(1984)}]{mateo+szkody84-1}
{Mateo}, M. \& {Szkody}, P. 1984, \aj, 89, 863

\bibitem[{{Mouchet} {et~al.}(1991){Mouchet}, {Bonnet-Bidaud}, {Buckley}, \&
  {Tuohy}}]{mouchetetal91-1}
{Mouchet}, M., {Bonnet-Bidaud}, J.~M., {Buckley}, D. A.~H., \& {Tuohy}, I.~R.
  1991, \aap, 250, 99

\bibitem[{{O'Donoghue} {et~al.}(2003){O'Donoghue}, {Koen}, {Kilkenny},
  {Stobie}, {Koester}, {Bessell}, {Hambly}, \&
  {MacGillivray}}]{odonoghueetal03-1}
{O'Donoghue}, D., {Koen}, C., {Kilkenny}, D., {Stobie}, R.~S., {Koester}, D.,
  {Bessell}, M.~S., {Hambly}, N., \& {MacGillivray}, H. 2003, \mnras, 345, 506

\bibitem[{{Paczy\'nski}(1983)}]{Pacz83}
{Paczy\'nski}, B. 1983, \apj, 267, 315

\bibitem[{{Paczynski} \& {Sienkiewicz}(1981)}]{PaczSien81}
{Paczynski}, B. \& {Sienkiewicz}, R. 1981, \apjl, 248, L27

\bibitem[{{Paczynski} \& {Sienkiewicz}(1983)}]{PaczSien83}
---. 1983, \apj, 268, 825

\bibitem[{{Patterson}(1984)}]{patterson84-1}
{Patterson}, J. 1984, \apjs, 54, 443

\bibitem[{{Piro} {et~al.}(2005){Piro}, {Arras}, \& {Bildsten}}]{piroetal05-1}
{Piro}, A.~L., {Arras}, P., \& {Bildsten}, L. 2005, \apj, 628, 401

\bibitem[{{Piro} \& {Bildsten}(2004)}]{piro+bildsten04-1}
{Piro}, A.~L. \& {Bildsten}, L. 2004, \apjl, 616, L155

\bibitem[{{Prialnik}(1986)}]{prialnik86-1}
{Prialnik}, D. 1986, \apj, 310, 222

\bibitem[{{Rappaport} {et~al.}(1982){Rappaport}, {Joss}, \&
  {Webbink}}]{Rappetal82}
{Rappaport}, S., {Joss}, P.~C., \& {Webbink}, R.~F. 1982, \apj, 254, 616

\bibitem[{{Rappaport} {et~al.}(1983){Rappaport}, {Verbunt}, \&
  {Joss}}]{Rappetal83}
{Rappaport}, S., {Verbunt}, F., \& {Joss}, P.~C. 1983, \apj, 275, 713

\bibitem[{{Reimers} \& {Hagen}(2000)}]{reimers+hagen00-1}
{Reimers}, D. \& {Hagen}, H.~J. 2000, \aap, 358, L45

\bibitem[{{Ribas}(2006)}]{Riba06}
{Ribas}, I. 2006, \apss, 304, 89

\bibitem[{{Ritter} \& {Kolb}(2003)}]{ritter+kolb03-1}
{Ritter}, H. \& {Kolb}, U. 2003, \aap, 404, 301

\bibitem[{{Ritter} {et~al.}(2000){Ritter}, {Zhang}, \& {Kolb}}]{ritteretal00-1}
{Ritter}, H., {Zhang}, Z.~., \& {Kolb}, U. 2000, \aap, 360, 959

\bibitem[{{Rodr{\'{\i}}guez-Gil} {et~al.}(2007){Rodr{\'{\i}}guez-Gil},
  {G{\"a}nsicke}, {Hagen}, {Araujo-Betancor}, {Aungwerojwit}, {Allende Prieto},
  {Boyd}, {Casares}, {Engels}, {Giannakis}, {Harlaftis}, {Kube}, {Lehto},
  {Mart{\'{\i}}nez-Pais}, {Schwarz}, {Skidmore}, {Staude}, \&
  {Torres}}]{rodriguez-giletal07-2}
{Rodr{\'{\i}}guez-Gil}, P., {G{\"a}nsicke}, B.~T., {Hagen}, H.-J.,
  {Araujo-Betancor}, S., {Aungwerojwit}, A., {Allende Prieto}, C., {Boyd}, D.,
  {Casares}, J., {Engels}, D., {Giannakis}, O., {Harlaftis}, E.~T., {Kube}, J.,
  {Lehto}, H., {Mart{\'{\i}}nez-Pais}, I.~G., {Schwarz}, R., {Skidmore}, W.,
  {Staude}, A., \& {Torres}, M.~A.~P. 2007, \mnras, 377, 1747

\bibitem[{{Rogers} {et~al.}(1996){Rogers}, {Swenson}, \&
  {Iglesias}}]{RogeSwenIgle96}
{Rogers}, F.~J., {Swenson}, F.~J., \& {Iglesias}, C.~A. 1996, \apj, 456, 902

\bibitem[{{Rosen} {et~al.}(2001){Rosen}, {Rainger}, {Burleigh}, {Mittaz},
  {Buckley}, {Sirk}, {Lieu}, {Howell}, \& {de Martino}}]{rosenetal01-1}
{Rosen}, S.~R., {Rainger}, J.~F., {Burleigh}, M.~R., {Mittaz}, J. P.~D.,
  {Buckley}, D. A.~H., {Sirk}, M.~M., {Lieu}, R., {Howell}, S.~B., \& {de
  Martino}, D. 2001, \mnras, 322, 631

\bibitem[{{Schmidt} {et~al.}(2005){Schmidt}, {Szkody}, {Vanlandingham},
  {Anderson}, {Barentine}, {Brewington}, {Hall}, {Harvanek}, {Kleinman},
  {Krzesinski}, {Long}, {Margon}, {Neilsen}, {Newman}, {Nitta}, {Schneider}, \&
  {Snedden}}]{schmidtetal05-1}
{Schmidt}, G.~D., {Szkody}, P., {Vanlandingham}, K.~M., {Anderson}, S.~F.,
  {Barentine}, J.~C., {Brewington}, H.~J., {Hall}, P.~B., {Harvanek}, M.,
  {Kleinman}, S.~J., {Krzesinski}, J., {Long}, D., {Margon}, B., {Neilsen},
  Jr., E.~H., {Newman}, P.~R., {Nitta}, A., {Schneider}, D.~P., \& {Snedden},
  S.~A. 2005, \apj, 630, 1037

\bibitem[{{Schmidt} {et~al.}(1986){Schmidt}, {West}, {Liebert}, {Green}, \&
  {Stockman}}]{schmidtetal86-2}
{Schmidt}, G.~D., {West}, S.~C., {Liebert}, J., {Green}, R.~F., \& {Stockman},
  H.~S. 1986, \apj, 309, 218

\bibitem[{{Schreiber} \& {G{\" a}nsicke}(2003)}]{schreiber+gaensicke03-1}
{Schreiber}, M.~R. \& {G{\" a}nsicke}, B.~T. 2003, \aap, 406, 305

\bibitem[{{Schwope} {et~al.}(2002){Schwope}, {Hambaryan}, {Schwarz}, {Kanbach},
  \& {G{\" a}nsicke}}]{schwopeetal02-1}
{Schwope}, A.~D., {Hambaryan}, V., {Schwarz}, R., {Kanbach}, G., \& {G{\"
  a}nsicke}, B.~T. 2002, \aap, 392, 541

\bibitem[{{Shafter} {et~al.}(1985){Shafter}, {Szkody}, {Liebert}, {Penning},
  {Bond}, \& {Grauer}}]{shafteretal85-1}
{Shafter}, A.~W., {Szkody}, P., {Liebert}, J., {Penning}, W.~R., {Bond}, H.~E.,
  \& {Grauer}, A.~D. 1985, \apj, 290, 707

\bibitem[{{Shara} {et~al.}(1986){Shara}, {Livio}, {Moffat}, \&
  {Orio}}]{1986ApJ...311..163S}
{Shara}, M.~M., {Livio}, M., {Moffat}, A.~F.~J., \& {Orio}, M. 1986, \apj, 311,
  163

\bibitem[{{Sion}(1991)}]{Sion91}
{Sion}, E.~M. 1991, \aj, 102, 295

\bibitem[{{Sion}(1995)}]{Sion95}
---. 1995, \apj, 438, 876

\bibitem[{{Sion}(1999)}]{Sion99}
---. 1999, \pasp, 111, 532

\bibitem[{{Sion} {et~al.}(2004){Sion}, {Cheng}, {Godon}, \&
  {Szkody}}]{sionetal04-1}
{Sion}, E.~M., {Cheng}, F., {Godon}, P., \& {Szkody}, P. 2004, \apj, in press

\bibitem[{{Sion} {et~al.}(1996){Sion}, {Cheng}, {Huang}, {Hubeny}, \&
  {Szkody}}]{sionetal96-1}
{Sion}, E.~M., {Cheng}, F., {Huang}, M., {Hubeny}, I., \& {Szkody}, P. 1996,
  \apjl, 471, L41

\bibitem[{{Sion} {et~al.}(1995{\natexlab{a}}){Sion}, {Cheng}, {Long}, {Szkody},
  {Gilliland}, {Huang}, \& {Hubeny}}]{sionetal95-2}
{Sion}, E.~M., {Cheng}, F.~H., {Long}, K.~S., {Szkody}, P., {Gilliland}, R.~L.,
  {Huang}, M., \& {Hubeny}, I. 1995{\natexlab{a}}, \apj, 439, 957

\bibitem[{{Sion} {et~al.}(1998){Sion}, {Cheng}, {Szkody}, {Sparks},
  {G\"ansicke}, {Huang}, \& {Mattei}}]{sionetal98-1}
{Sion}, E.~M., {Cheng}, F.~H., {Szkody}, P., {Sparks}, W., {G\"ansicke}, B.,
  {Huang}, M., \& {Mattei}, J. 1998, \apj, 496, 449

\bibitem[{{Sion} {et~al.}(1990){Sion}, {Leckenby}, \& {Szkody}}]{sionetal90-1}
{Sion}, E.~M., {Leckenby}, H.~J., \& {Szkody}, P. 1990, \apjl, 364, L41

\bibitem[{{Sion} {et~al.}(2003){Sion}, {Szkody}, {Cheng}, {G{\"a}nsicke}, \&
  {Howell}}]{sionetal03-1}
{Sion}, E.~M., {Szkody}, P., {Cheng}, F., {G{\"a}nsicke}, B.~T., \& {Howell},
  S.~B. 2003, \apj, 583, 907

\bibitem[{{Sion} {et~al.}(1995{\natexlab{b}}){Sion}, {Szkody}, {Cheng}, \&
  {Huang}}]{sionetal95-3}
{Sion}, E.~M., {Szkody}, P., {Cheng}, F., \& {Huang}, M. 1995{\natexlab{b}},
  \apjl, 444, L97

\bibitem[{{Sion} {et~al.}(2001){Sion}, {Szkody}, {G\"ansicke}, {Cheng},
  {LaDous}, \& {Hassall}}]{sionetal01-1}
{Sion}, E.~M., {Szkody}, P., {G\"ansicke}, B., {Cheng}, F.~H., {LaDous}, C., \&
  {Hassall}, B. 2001, \apj, 555, 834

\bibitem[{{Sion} \& {Urban}(2002)}]{sion+urban02-1}
{Sion}, E.~M. \& {Urban}, J. 2002, \apj, 572, 456

\bibitem[{{Smith} {et~al.}(2006){Smith}, {Haswell}, \& {Hynes}}]{smithetal06-1}
{Smith}, A.~J., {Haswell}, C.~A., \& {Hynes}, R.~I. 2006, \mnras, 369, 1537

\bibitem[{{Smith} \& {Dhillon}(1998)}]{smith+dhillon98-1}
{Smith}, D.~A. \& {Dhillon}, V.~S. 1998, \mnras, 301, 767

\bibitem[{{Southworth} {et~al.}(2006){Southworth}, {G{\"a}nsicke}, {Marsh}, {de
  Martino}, {Hakala}, {Littlefair}, {Rodr{\'{\i}}guez-Gil}, \&
  {Szkody}}]{southworthetal06-1}
{Southworth}, J., {G{\"a}nsicke}, B.~T., {Marsh}, T.~R., {de Martino}, D.,
  {Hakala}, P., {Littlefair}, S., {Rodr{\'{\i}}guez-Gil}, P., \& {Szkody}, P.
  2006, \mnras, 373, 687

\bibitem[{{Spruit} \& {Ritter}(1983)}]{SpruRitt83}
{Spruit}, H.~C. \& {Ritter}, H. 1983, \aap, 124, 267

\bibitem[{{Steeghs} {et~al.}(2007){Steeghs}, {Howell}, {Knigge},
  {G{\"a}nsicke}, {Sion}, \& {Welsh}}]{steeghsetal07-1}
{Steeghs}, D., {Howell}, S.~B., {Knigge}, C., {G{\"a}nsicke}, B.~T., {Sion},
  E.~M., \& {Welsh}, W.~F. 2007, \apj, 667, 442

\bibitem[{{Steeghs} {et~al.}(2001){Steeghs}, {Marsh}, {Knigge}, {Maxted},
  {Kuulkers}, \& {Skidmore}}]{steeghsetal01-2}
{Steeghs}, D., {Marsh}, T., {Knigge}, C., {Maxted}, P.~F.~L., {Kuulkers}, E.,
  \& {Skidmore}, W. 2001, \apjl, 562, L145

\bibitem[{{Stockman} {et~al.}(1994){Stockman}, {Schmidt}, {Liebert}, \&
  {Holberg}}]{stockmanetal94-1}
{Stockman}, H.~S., {Schmidt}, G.~D., {Liebert}, J., \& {Holberg}, J.~B. 1994,
  \apj, 430, 323

\bibitem[{{Szkody} {et~al.}(2003{\natexlab{a}}){Szkody}, {Anderson}, {Schmidt},
  {Hall}, {Margon}, {Miceli}, {SubbaRao}, {Frith}, {Harris}, {Hawley},
  {Lawton}, {Covarrubias}, {Covey}, {Fan}, {Murphy}, {Narayanan}, {Raymond},
  {Rest}, {Strauss}, {Stubbs}, {Turner}, {Voges}, {Bauer}, {Brinkmann},
  {Knapp}, \& {Schneider}}]{szkodyetal03-3}
{Szkody}, P., {Anderson}, S.~F., {Schmidt}, G., {Hall}, P.~B., {Margon}, B.,
  {Miceli}, A., {SubbaRao}, M., {Frith}, J., {Harris}, H., {Hawley}, S.,
  {Lawton}, B., {Covarrubias}, R., {Covey}, K., {Fan}, X., {Murphy}, T.,
  {Narayanan}, V., {Raymond}, S., {Rest}, A., {Strauss}, M.~A., {Stubbs}, C.,
  {Turner}, E., {Voges}, W., {Bauer}, A., {Brinkmann}, J., {Knapp}, G.~R., \&
  {Schneider}, D.~P. 2003{\natexlab{a}}, \apj, 583, 902

\bibitem[{{Szkody} {et~al.}(2000{\natexlab{a}}){Szkody}, {Desai}, {Burdullis},
  {Hoard}, {Fried}, {Garnavich}, \& {G\"ansicke}}]{szkodyetal00-3}
{Szkody}, P., {Desai}, V., {Burdullis}, T., {Hoard}, D.~W., {Fried}, R.,
  {Garnavich}, P., \& {G\"ansicke}, B. 2000{\natexlab{a}}, \apj, 540, 983

\bibitem[{{Szkody} {et~al.}(2000{\natexlab{b}}){Szkody}, {Desai}, \&
  {Hoard}}]{Szkoetal00}
{Szkody}, P., {Desai}, V., \& {Hoard}, D.~W. 2000{\natexlab{b}}, \aj, 119, 365

\bibitem[{{Szkody} {et~al.}(2002{\natexlab{a}}){Szkody}, {G{\" a}nsicke},
  {Howell}, \& {Sion}}]{szkodyetal02-4}
{Szkody}, P., {G{\" a}nsicke}, B.~T., {Howell}, S.~B., \& {Sion}, E.~M.
  2002{\natexlab{a}}, \apjl, 575, L79

\bibitem[{{Szkody} {et~al.}(2002{\natexlab{b}}){Szkody}, {G{\" a}nsicke},
  {Sion}, \& {Howell}}]{szkodyetal02-3}
{Szkody}, P., {G{\" a}nsicke}, B.~T., {Sion}, E.~M., \& {Howell}, S.~B.
  2002{\natexlab{b}}, \apj, 574, 950

\bibitem[{{Szkody} {et~al.}(2003{\natexlab{b}}){Szkody}, {G{\" a}nsicke},
  {Sion}, {Howell}, \& {Cheng}}]{szkodyetal03-1}
{Szkody}, P., {G{\" a}nsicke}, B.~T., {Sion}, E.~M., {Howell}, S.~B., \&
  {Cheng}, F.~H. 2003{\natexlab{b}}, \aj, 126, 1451

\bibitem[{{Szkody} {et~al.}(2006){Szkody}, {Harrison}, {Plotkin}, {Howell},
  {Seibert}, \& {Bianchi}}]{szkodyetal06-2}
{Szkody}, P., {Harrison}, T.~E., {Plotkin}, R.~M., {Howell}, S.~B., {Seibert},
  M., \& {Bianchi}, L. 2006, \apjl, 646, L147

\bibitem[{{Szkody} {et~al.}(2007){Szkody}, {Mukadam}, {G{\"a}nsicke}, {Woudt},
  {Solheim}, {Nitta}, {Sion}, {Warner}, {Sahu}, {Prabhu}, \&
  {Henden}}]{szkodyetal07}
{Szkody}, P., {Mukadam}, A., {G{\"a}nsicke}, B.~T., {Woudt}, P.~A., {Solheim},
  J.-E., {Nitta}, A., {Sion}, E.~M., {Warner}, B., {Sahu}, D.~K., {Prabhu}, T.,
  \& {Henden}, A. 2007, \apj, 658, 1188

\bibitem[{{Thorstensen}(2003)}]{thorstensen03-1}
{Thorstensen}, J.~R. 2003, \aj, 126, 3017

\bibitem[{{Townsley} \& {Bildsten}(2003)}]{TownBild03}
{Townsley}, D.~M. \& {Bildsten}, L. 2003, \apjl, 596, L227

\bibitem[{{Townsley} \& {Bildsten}(2004)}]{TownBild04}
---. 2004, \apj, 600, 390, {TB}

\bibitem[{{Townsley} \& {Bildsten}(2005)}]{TownBild05}
---. 2005, \apj, 628, 395

\bibitem[{{Urban} \& {Sion}(2006)}]{urban+sion06-1}
{Urban}, J.~A. \& {Sion}, E.~M. 2006, \apj, 642, 1029

\bibitem[{{Verbunt} {et~al.}(1997){Verbunt}, {Bunk}, {Ritter}, \&
  {Pfeffermann}}]{verbuntetal97-1}
{Verbunt}, F., {Bunk}, W.~H., {Ritter}, H., \& {Pfeffermann}, E. 1997, \aap,
  327, 602

\bibitem[{{Verbunt} \& {Zwaan}(1981)}]{VerbZwaa81}
{Verbunt}, F. \& {Zwaan}, C. 1981, \aap, 100, L7

\bibitem[{{Vogel} {et~al.}(2007){Vogel}, {Schwope}, \&
  {G{\"a}nsicke}}]{vogeletal07-1}
{Vogel}, J., {Schwope}, A.~D., \& {G{\"a}nsicke}, B.~T. 2007, \aap, 464, 647

\bibitem[{{Warner}(1995)}]{Warn95}
{Warner}, B. 1995, {Cataclysmic Variable Stars} (Cambridge: Cambridge Univ.\
  Press)

\bibitem[{{Webbink} \& {Wickramasinghe}(2002)}]{webbink+wickramasinghe02-1}
{Webbink}, R.~F. \& {Wickramasinghe}, D.~T. 2002, \mnras, 335, 1

\bibitem[{{Webbink} \& {Wickramasinghe}(2005)}]{webbink+wickramasinghe05-1}
{Webbink}, R.~F. \& {Wickramasinghe}, D.~T. 2005, in The Astrophysics of
  Cataclysmic Variables and Related Objects, ed. J.-M. {Hameury} \& J.-P.
  {Lasota} (ASP Conf. Ser. 330), 137--146

\bibitem[{{Wickramasinghe} \& {Wu}(1994)}]{wickramasinghe+wu94-1}
{Wickramasinghe}, D.~T. \& {Wu}, K. 1994, \mnras, 266, L1

\bibitem[{{Winter} \& {Sion}(2003)}]{WintSion03}
{Winter}, L. \& {Sion}, E.~M. 2003, \apj, 582, 352

\bibitem[{{Wood} \& {Horne}(1990)}]{wood+horne90-1}
{Wood}, J.~H. \& {Horne}, K. 1990, \mnras, 242, 606

\bibitem[{{Wood} {et~al.}(1992){Wood}, {Horne}, \& {Vennes}}]{woodetal92-1}
{Wood}, J.~H., {Horne}, K., \& {Vennes}, S. 1992, \apj, 385, 294

\bibitem[{{Wood} {et~al.}(1995){Wood}, {Naylor}, {Hassall}, \&
  {Ramseyer}}]{woodetal95-2}
{Wood}, J.~H., {Naylor}, T., {Hassall}, B.~J.~M., \& {Ramseyer}, T.~F. 1995,
  \mnras, 273, 772

\bibitem[{{Wu} {et~al.}(1995){Wu}, {Wickramasinghe}, \& {Warner}}]{wuetal95-3}
{Wu}, K., {Wickramasinghe}, D.~T., \& {Warner}, B. 1995, Publications of the
  Astronomical Society of Australia, 12, 60

\end{thebibliography}
%\bibliography{aamnem99,aabib}

\end{document}